\documentclass[useAMS,usenatbib]{mn2e}

\usepackage{epsfig}
\usepackage{myaasmacros}
\usepackage{enumerate}
\usepackage{amsmath}
\usepackage{amsfonts}

\def\lesssim{\lower.5ex\hbox{$\; \buildrel < \over \sim \;$}}
\def\gtrsim{\lower.5ex\hbox{$\; \buildrel > \over \sim \;$}}

\title[Origin of the anti-hierarchical growth of black holes]
{Origin of the anti-hierarchical growth of black holes}
\author[Hirschmann et al.]{Michaela Hirschmann$^{1,2}$\thanks{E-mail:
mhirsch@oats.inaf.it}, Rachel S. Somerville$^{3}$, Thorsten Naab$^{4}$,
Andreas Burkert$^{2}$\\
$^{1}$INAF - Astronomical Observatory of Trieste, via G.B. Tiepolo 11,
I-34143 Trieste, Italy\\
$^{2}$Universit\"ats Sternwarte M\"unchen, Scheinerstr.1, D-81679 M\"unchen, Germany\\
$^{3}$Department of Physics and Astronomy, Rutgers, The State University of New Jersey, 136 Frelinghuysen Rd, Piscataway, NJ \\
$^{4}$Max-Plank-Institute f\"ur Astrophysik,
Karl-Schwarzschild Strasse 1,
D-85740 Garching, Germany}

\begin{document}

\date{Accepted ???. Received ??? in original form ???}

\pagerange{\pageref{firstpage}--\pageref{lastpage}} \pubyear{2002}

\maketitle

\label{firstpage}

\begin{abstract}

Observational studies have revealed a ``downsizing'' trend in black
hole (BH) growth: the number densities of luminous AGN peak at higher
redshifts than those of faint AGN. This would seem to imply that
massive black holes formed before low mass black holes, in apparent
contradiction to hierarchical clustering scenarios. We investigate
whether this observed ``downsizing'' in BH growth is reproduced in a
semi-analytic model for the formation and evolution of galaxies and
black holes, set within the hierarchical paradigm for structure
formation (Somerville et al. 2008; S08). In this model, black holes
evolve from light seeds ($\sim 100 M_{\odot}$) and their growth is
merger-driven. The original S08 model (baseline model) reproduces the
number density of AGN at intermediate redshifts and luminosities, but
underproduces luminous AGN at very high redshift ($z>3$) and
overproduces them at low redshift ($z<1$). In addition, the baseline
model underproduces low-luminosity AGN at low redshift ($z<1$). In
order to solve these problems, we consider several modifications to
the physical processes in the model: (1) a `heavy' black hole seeding
scenario (2) a sub-Eddington accretion rate ceiling that depends on the
cold gas fraction, and (3) an additional black hole accretion mode due
to disk instabilities. With these three modifications, the models can
explain the observed downsizing, successfully reproduce the bolometric
AGN luminosity function and simultaneously reproduce galaxy and black
hole properties in the local Universe. We also perform a comparison
with the observed soft and hard X-ray luminosity functions of AGN,
including an empirical correction for torus-level obscuration, and
reach similar conclusions. 
Our best-fit model suggests a scenario in which disk instabilities are
the main driver for moderately luminous Seyfert galaxies at low
redshift, while major mergers are the main trigger for luminous AGN. 
\end{abstract}

\begin{keywords}
keywords
\end{keywords}

\section{Introduction}\label{intro}

It has been known since early optical quasar surveys that the co-moving
number density of luminous quasars has a pronounced peak at a redshift
around $z=2$-2.5 (\citealp{Schmidt83, Boyle88, Hewett94, Boyle00,
  Warren94, Schmidt95}), with a fairly steep decline at higher and
lower redshift. More recently, it was discovered that very massive BH
($10^9 M_{\odot}$) appear to exist already at $z\sim 6$ and higher
\citep{Fan00, Fan01, Mortlock11}, but they are extremely rare, in
accordance with this trend. Recent progress in detecting faint and
obscured AGN has been achieved by analysing data from X-ray surveys
(XMM-Newton, Chandra, ROSAT, ASCA, e.g. \citealp{Miyaji00, LaFranca02,
  Cowie03, Fiore03, Barger03, Ueda03, Hasinger05, Barger05, Sazonov04,
  Nandra05, Ebrero09,Aird10,Fiore12}). All of these studies in the
hard and soft X-ray range find that the cosmic evolution of AGN is
strongly dependent on the AGN luminosity: the number density of
successively less luminous AGN peaks at lower redshifts, with the
lowest luminosity AGN showing a basically constant number density.
Making the simplified assumption that AGN luminosity is proportional
to BH mass (as we would expect if black holes are accreting at the
Eddington rate, $L \propto M_{\bullet}$) would imply that very massive
black holes seem to be already in place at very early times, whereas
less massive black holes seem to evolve predominantly at lower
redshifts. This behavior is called `downsizing' or `anti-hierarchical'
growth of black holes. The downsizing trend is also seen in the
optical (\citealp{Cristiani04, Croom04, Fan04, Hunt04, Richards06,
  Wolf03}) and the NIR (e.g. \citealp{Matute06}).  

On the face of it, this observational result seems to be in conflict
with the expectations within the currently favored hierarchical
structure formation paradigm, such as those based on the Cold Dark
Matter (CDM) model (\citealp{Peebles65,White78,Blumenthal85}). In
this framework, low mass halos form first and more massive halos grow
over time via subsequent merging and smooth accretion. However it is
now well known that the evolutionary history of observable
\emph{galaxies} also follows an ``anti-hierarchical'' or downsizing
behavior, with several independent observational indicators suggesting
that more massive galaxies formed their stars and had their star
formation ``quenched'' earlier than low-mass galaxies, which continue
forming stars to the present day (an overview of these observations is
given in \citealt{Fontanot09}).

Present-day spheroidal galaxies host supermassive black holes at their
centers (\citealp{Magorrian98, Genzel99}) and strong correlations have
been found between black hole masses and properties of their host
galaxies (\citealp{Ferrarese00, Gebhardt00, Tremaine02, Haering04,
  Graham12}). 
This can be interpreted as evidence for co-evolution between the host
galaxies and their black holes, but some observations of BH growth and
SF in individual objects appear to contradict the picture of simple
one-to-one co-evolution over time \citep{Mullaney12}. Thus the
details of BH and galaxy co-evolution remain unclear. 

During their lifetime, black holes are assumed to undergo several
episodes of significant gas accretion, during which this accretion
powers luminous quasars or active galactic nuclei (AGN)
(\citealp{Salpeter64, Zeldovich64, Lynden69}). By estimating the total
energy radiated by AGN over their whole lifetime, it can be shown that
nearly all the mass seen in dormant black holes today can be
accumulated during the periods of observed bright AGN activity
(\citealp{Soltan82}). 
This implies that there is not a great deal of room for ``dark'' or
obscured accretion.

A large number of works have explored the predictions of the
$\Lambda$CDM model for the formation and evolution of supermassive
black holes and AGN, with varying levels of complexity. Several works
made predictions based on nearly purely analytic models
(\citealp{Efstathiou88, Haehnelt93, Haiman98}) and on semi-empirical models
  (\citealp{Shankar09, Shankar10a, Shankar12}) and a
  large number of studies have been published based on semi-analytic
  models of galaxy formation within which mechanisms for black hole
  formation and evolution have been included (e.g. \citealp{Kauffmann00,
    Volonteri03, Granato04, Menci04, Bromley04, Croton06, Bower06,
    Marulli08, Somerville08}). 
Recently, numerical hydrodynamic simulations have also included black
hole growth and AGN feedback using ``sub-grid'' recipes
(\citealp{Springel05b, Hopkins06, DiMatteo05, Robertson06c, Li07,
  Sijacki07, Johansson09_1, McCarthy10, McCarthy11, Degraf11,
  DiMatteo12, Choi12}).  
\citet{Hopkins08a} presented predictions based on
``semi-empirical'' models in which galaxy properties were taken from
observations and the relationship between galaxy properties and AGN
luminosity was based on the results of a large suite of hydrodynamic
merger simulations.

These models differ in many of the details of how BH formation and
growth are implemented, but there seems to be a broad consensus on
several points. First, self-regulated black hole growth, perhaps via
radiation pressure driven winds (\citealp{DiMatteo05, King05,
  Murray05, Robertson06b}) is a widely adopted means of obtaining the
observed tight relationship between BH mass and spheroid mass
(although some models simply assume this relationship without invoking
a physical mechanism). Second, one must invoke a physical mechanism
that can feed large amounts of gas onto the central BH within a short
time. Galaxy-galaxy mergers are a popular (though not universally
adopted) way to remove angular momentum from the gas and efficiently
drive it to the center of the galaxy. Third, some mechanism must be
adopted that reduces or stops gas cooling and accretion in massive
dark matter halos. Many models assume that low levels of accretion
onto SMBH can produce radio jets that heat the surrounding hot halo
\citep{Croton06,Bower06,Somerville08} thereby inhibiting cooling
flows, preventing over-massive galaxies from forming and quenching
star formation in massive galaxies, leading to more observationally
consistent color-magnitude or SFR-stellar mass relationships
\citep{kimm:09}.

However, several aspects of the physics of black hole growth and
formation remain poorly understood. First, there are active ongoing
debates about when and how seed black holes form: either via a direct
collapse of cold gas clouds leading to massive seeds of $\sim
10^4$-$10^5 M_{\odot}$
(\citealp{Loeb94,Koushiappas04,Volonteri11,Bellovary11}) or via
stellar remnants from Pop III stars, resulting in low mass ($\sim 100
M_{\odot}$) black hole seeds (\citealp{Madau01,Heger02}). An
alternative possibility is direct seed formation in a merger event as
seen in the numerical simulations of \citet{Mayer10}. While the
seeding mechanisms do not alter the AGN and black hole population at
low redshift (as gas accretion during the evolution exceeds the seed
black hole masses by many orders of magnitude), at high redshift the
choice of the seeding model strongly influences the black hole
formation and is highly relevant for understanding the observed
population of luminous high redshift quasars.

Another matter of vigorous debate is the process or processes that
trigger and regulate accretion onto the central SMBH.  As noted above,
the most luminous observed quasars require accretion rates such that
$\sim 10^8$--$10^9 M_{\odot}$ of gas must be funneled onto the black
hole over a timescale of $<10^{8.5}$ yr, i.e. nearly the whole gas
content of a good-sized galaxy must be fed onto the black hole in
roughly a dynamical time \citep[see discussion in][and references
  therein]{Hopkins08a}.
Mergers appear to be a physically well-motivated candidate for
producing this dramatic effect, and semi-empirical calculations have
shown that there is a statistical consistency between the observed
merger rate and the observed AGN duty cycle and luminosity function
(\citealp{Hopkins06, Hopkins08a}), 
suggesting that it is at
least possible that these processes are causally linked. However, the
observational situation remains murky. Observational studies have
repeatedly failed to find evidence for a statistically significant
enhancement of merger-related signatures, such as close pairs or
morphological disturbance, in AGN hosts up to $z\sim 1$
(\citealp{Cisternas10, Georgakakis09, Pierce07, Grogin05, Li08,
  Ellison08}).
But a recent study by \citet{Ellison11} 
does find a significant
enhancement of AGN activity in close pairs, and discusses reasons that
previous studies based on similar data have not found a signal.
The absence of enhanced merger signatures in morphological studies has
recently been shown to persist in X-ray selected AGN up to $z\sim 2.5$
\citep{Schawinski11,Kocevski12}. Recent work suggests, though, that the
fraction of hosts with morphological disturbances may be higher in
obscured AGN, many of which are missing from X-ray selected surveys
(S. Juneau, private communication).

The goal of this work is to explore the the interplay of different
physical processes that determine the masses of seed black holes, the
triggering of AGN activity and the efficiency of gas accretion during
the active phases of black holes, with the aim of understanding the
physical origin of the observed downsizing trend in black hole
growth. We follow a semi-analytic approach, which has been shown to
successfully reproduce many observed galaxy population properties
(\citealp{Somerville08}, \citealp{Somerville11}) and includes a model
for the merger triggered formation and evolution of black holes (see
Section \ref{model} for details). In this paper we present three major
modifications to the baseline model published in S08:
\begin{itemize}
\item{gas fraction dependent Eddington ratios (accretion efficiency)}
\item{AGN activity triggered by disk instabilities} 
\item{`heavy' black hole seeds} 
\end{itemize}
to the existing model. We find that with these three modifications,
our model does reproduce the observed downsizing behavior.

In Section \ref{previous} we briefly summarize the results from
previous studies. Section \ref{model} gives a brief overview of the
semi-analytic model used in our study and describes the different
modifications for black hole growth we are considering. In Sections
\ref{obs} and \ref{BHpropevol} we present some properties of
present-day and high-redshift galaxies and their black holes, which
are compared to observations. The AGN number density evolution is
studied in Section \ref{Numdens}, considering the influence of the
different modifications concerning black hole growth. Section
\ref{AGNlum} presents a comparison of the evolution of the observed
bolometric and hard and soft X-ray luminosity function with the model
output. In Sections \ref{Eddratioevol} and \ref{BHlumevol}, we discuss
the evolution of the Eddington ratio distributions and the evolution
of the black hole-AGN luminosity plane. Finally, in Section
\ref{downdis}, we summarize and discuss our main results.

\section{Previous studies}\label{previous} 

A number of studies have investigated the observed anti-hierarchical
trend of BH activity using the \textsc{`Galform'} (Durham)
(\citealp{Bower06}), the \textsc{`Munich'} (\citealp{DeLucia07}) or
the \textsc{`Morgana'} semi-analytic models (\citealp{Monaco05,
  Fontanot06}). The first two models are applied to the dark matter
merger trees of the Millennium simulation, while the latter uses
merger trees from the \textsc{Pinocchio} method.  All models
distinguish between black hole accretion in the bright quasar mode and
the low-Eddington ratio radio-mode.  In the \textsc{`Galform'} model,
the quasar mode is triggered by merger events \textit{and} disk
instabilities, but the mass growth of black holes is dominated by
accretion due to disk instabilities. In the \textsc{`Munich'}
model the quasar mode is assumed to be triggered \textit{only} by
merger events. 
The \textsc{`Morgana'} model assumes that any low angular momentum gas
within the ``bulge'' gas reservoir is available to accrete onto the
central BH. Gas may be transferred to the ``low-$J$'' reservoir either
by mergers or disk instabilities. In this model, black holes grow at a
rate determined by the viscosity of the accretion disk and the mass of
gas in the low-$J$ reservoir. In the \textsc{`Galform'}-model the
black hole accretion rate is proportional to the star formation rate
during a starburst (triggered either by a merger or disk instability),
whereas in the \textsc{`Munich'}-model the accretion rate is dependent
on the cold gas content in the galaxy, the galaxy circular velocity,
and the mass ratio of the triggering merger.  
While these recipes differ in detail, they are both proportional to
the gas content and rougly inversely proportional to $(1+V^{-2}_c)$,
where $V_c$ is the circular velocity of the bulge component in the
case of \textsc{`Galform'} and the halo virial velocity in the
\textsc{`Munich'}-model, and both lead to a black hole-bulge mass
relation at $z=0$ that is consistent with the observed one. However,
it appears that the history of black hole accretion is significantly
different in the two models. In the \textsc{`Morgana'}-model, black
holes grow at a rate determined by the viscosity of the accretion disk
and the mass of gas in the low-$J$ reservoir.

\citet{Fontanot06} claim that their \textsc{`Morgana'}-model can
reproduce downsizing, which in their model is caused by stellar
kinetic feedback that arises in star-forming bulges leading to a
removal of cold gas in small elliptical galaxies (reduction of the
number of faint AGN at high redshift). To obtain a good match to the
number density of bright quasars they require quasar-triggered
galactic winds, which self-limit the accretion onto black holes.

\citet{Malbon07} found, using the \textsc{`Galform'} semi-analytic
code, that the direct accretion of cold gas during starbursts is an
important growth mechanism for lower mass black holes and and for all
black holes at high redshift. The assembly of pre-existing black hole
mass into larger units via merging dominates the growth of more
massive black holes at low redshift. Therefore, they claim that as
redshift decreases, progressively less massive black holes have the
highest growth rates, in agreement with downsizing. Their model output
reproduces the evolution of the optical luminosity function of
quasars, however, they do not show a quantitative comparison for the
X-ray and/or bolometric AGN luminosity.

\citet{Fanidakis10} have constructed a model based on the
\textsc{`Galform'}-model framework, but with different recipes for
black hole growth and AGN feedback. They present a quantitative
comparison of their model output to the observed quasar luminosity
function at different redshifts. 
The previous \textsc{`Galform'}-based SAMs associated radiatively
efficient BH accretion with ``cold mode'' accretion (merger or
disc-instability triggered) and radiatively inefficient accretion with
``hot mode'' accretion (accretion from a quasi-hydrostatic hot gas
halo). In contrast, \citet{Fanidakis10} associate accretion at low
accretion rates (less than about one percent of the Eddington rate),
regardless of its origin, with radiatively inefficient advection
dominated accretion flows (ADAF), and assume that accretion at higher
rates will be radiatively efficient. At high redshift,
\citet{Fanidakis10} do not Eddington-limit the accretion rates, but
allow super-Eddington accretion, which is responsible for very
luminous AGN at high redshifts. The high number density of low
luminosity AGN at low redshift can be explained by the luminosity
produced via the ADAF mode. They attribute the observed downsizing
trend mainly to dust obscuration of low luminosity AGN at high
redshift. 

\citet{Marulli08} investigated different parameterizations of the
quasar light curves in the \textsc{`Munich'}-model, and found as
expected, that the lightcurve parameterization has a large effect on
the number density of faint AGN as a function of redshift. They found
the best results with a lightcurve model that includes an Eddington
growth phase followed by a power-law decline phase as suggested by
\citet{Hopkins06}. However, for all adopted lightcurve models they
found that the previously published Munich model underpredicts the
number density of luminous quasars at high redshift ($z>1$). Finally,
in a follow-up study by \citet{Bonoli09}, the BH accretion efficiency
was assumed to be a function of redshift as well as gas content and
merger mass ratio. They obtained improved results, but still were not
able to reproduce the luminosity function of observed AGN over the
full range in redshift and luminosity. 

\section{The semi-analytic model}\label{model} 

The semi-analytic model used in this study is presented in S08 and we
refer the reader to this paper for details. The galaxy formation model
is based on dark matter merger trees generated by the extended
Press-Schechter formalism. The evolution of baryons within these dark
matter halos is modeled using prescriptions for gas cooling,
re-ionization, star formation, supernova feedback, metal evolution,
black hole growth and AGN feedback. Here we focus on the mechanism
describing the formation and evolution of black holes.  \textit{Each}
top-level dark matter halo is seeded with a $100\ M_{\odot}$ black
hole in its center, which can grow by two mechanisms: through cold gas
accretion during the `bright' quasar mode and through accretion of gas
from the hot halo via a cooling flow during the low-Eddington ratio
and radiatively inefficient radio mode. The quasar mode is assumed to
produce momentum-driven winds, which are modeled using the analytic
scaling derived and calibrated from binary hydrodynamic merger
simulations (\citealp{Robertson06a, Cox06a, Robertson06c,
  Hopkins07a, Robertson06b}).  

The radio mode can only occur when a hot quasi-hydrostatic halo is
present, which is assumed to be the case when the cooling shock is
predicted to be within the virial radius ($r_{\mathrm{cool}} <
r_{\mathrm{vir}}$). During this phase, black holes are fuelled by
Bondi-Hoyle-accretion (\citealp{Bondi52}), with the isothermal
cooling-flow solution from \citet{Nulsen00}. The growth in the radio
mode is also associated with an efficient production of radio jets
that results in an energy injection into the intracluster medium
(ICM). Therefore, we assume that the energy arising from the accretion
onto the black hole couples to and heats the gas in the surrounding
hot halo (Radio mode feedback).

\subsection{Standard accretion model}\label{FID}

The quasar phase is triggered by galaxy merger events with a mass
ratio of $\mu > 0.1$, where $\mu$ is the mass ratio of the baryonic
components and the dark matter within the central part of the galaxy
(see S08 for the precise definition). The lower limit is motivated by
binary hydrodynamic merger simulations. Whenever the two progenitor
galaxies merge, their black holes are assumed to also merge and form a
single black hole whose mass is the sum of the progenitor BH's
masses. The model for gas accretion onto the black hole is motivated
by the analysis of gas inflow rates onto the nuclear regions from
idealized disk merger simulations (\citealp{Springel05b, Robertson06c,
  Cox06, Hopkins06, Hopkins07b}). During the merger, the BH is assumed
to grow rapidly with accretion rates near the Eddington limit. This
rapid accretion continues until the energy being deposited into the
ISM in the central region of the galaxy is sufficient to significantly
offset and eventually halt accretion via a pressure-driven
outflow. During this ``blow-out'' phase, the accretion rate declines
gradually until the nuclear fuel is exhausted. 

Based on the merger simulations, the final black hole mass
$M_{\bullet,\mathrm{final}}$ at the end of the blow-out phase is
assumed to be related to the mass of the spheroidal component after
the merger:
\begin{eqnarray}\label{mbhfinal} 
M_{\bullet,\mathrm{final}} = f_{\mathrm{BH,final}} \times 0.158 \left(
  \frac{M_{\mathrm{sph}}}{100 M_{\odot}} \right)^{1.12} \times
\Gamma(z).
\end{eqnarray}
Here, $M_{\mathrm{sph}}$ is the final spheroid mass after the merger,
$f_{\mathrm{BH,final}}$ is an adjustable parameter and $\Gamma(z)$
describes the evolution of the black hole-bulge mass relation with
time (\citealp{Hopkins06a}). Following the merger simulations, a
Gaussian distributed scatter with a value of $\sigma_{\bullet} =
0.3\ \mathrm{dex}$ is additionally applied to the accreted gas mass,
representing stochasticity due to e.g. the properties of the
orbit. When the black hole mass has reached its final mass value, the quasar
mode is switched off. During the quasar phase, the light curve models
describe two different growth regimes: an Eddington-limited and a
power-law decline phase of accretion. In the first regime, the black
hole accretes at the Eddington limit until it reaches a critical black
hole mass $M_{\bullet,\mathrm{crit}}$:
\begin{eqnarray}\label{Mcrit}
M_{\bullet,\mathrm{crit}} = f_{\mathrm{BH,crit}} \times 1.07
\left( M_{\bullet,\mathrm{final}}\right)^{1.1}.
\end{eqnarray}
Here, the parameter $f_{\mathrm{BH,crit}}$ is set according to the
merger simulations, and determines how much of the black hole growth
occurs in the Eddington-limited versus power-law decline phase. The
growth of the black hole during the first regime can be modeled by an
exponential increase of mass:
\begin{eqnarray}\label{expmass}
 M_{\bullet,\mathrm{new}}(t) = M_{\bullet}
 \exp \left(\frac{1-\epsilon}{\epsilon} f_{\mathrm{edd}}
   \frac{t}{t_{\mathrm{salp}}} \right),
\end{eqnarray}
where $\epsilon = 0.1$ is the efficiency of the conversion of rest
mass to energy, and $t_{\mathrm{salp}} \approx 0.45\ \mathrm{Gyr}$ is
the Salpeter timescale. No strong observational constraints are
available for $\epsilon$ and if or how it evolves with
redshift. However, some observations at $z=0$ indicate that $0.04 <
\epsilon < 0.16$ (\citealp{Marconi04}). For simplicity we take a
constant mean value of $\epsilon = 0.1$ at all redshifts, which is a
very standard assumption. The parameter $f_{\mathrm{Edd}}$ in
Eq. \ref{expmass} is the maximum accretion rate, which is defined by
the ratio of bolometric luminosity to the Eddington luminosity:
\begin{eqnarray}\label{eddratio}
f_{\mathrm{Edd}} = L_{\mathrm{bol}}/L_{\mathrm{Edd}}.
\end{eqnarray}
The Eddington luminosity $L_{\mathrm{Edd}}$ (assuming a hydrostatic
equilibrium between the inward gravitational force and the outward
radiation pressure) is given by:
\begin{eqnarray}\label{eddlum}
L_{\mathrm{Edd}} = \frac{4\pi G M_{\bullet} m_p\, c}{\sigma_T} = 1.4
\times 10^{46} \left( \frac{M_{\bullet}}{10^8M_{\odot}}\right)
\mathrm{erg/s},
\end{eqnarray}
where $\sigma_T$ is the Thomson cross section for an electron and
$m_p$ the mass of a proton. 
Combining eq. \ref{expmass}, \ref{eddratio} and \ref{eddlum} and the
relation $L_{\mathrm{bol}} = \epsilon/(1-\epsilon) \dot{M} c^2$, the
corresponding accretion rate in the first regime can be calculated by: 
\begin{eqnarray}
\dot{M}_{\bullet,I}(t) = 1.26 \times
10^{38} \mathrm{erg/s}\ \frac{1-\epsilon}{\epsilon}
\frac{f_{\mathrm{edd}}}{c^2} M_{\bullet,\mathrm{new}}(t) .
\end{eqnarray}
Note that in the baseline model, the maximum accretion rate is assumed
to equal the Eddington rate, i.e. $f_{\mathrm{Edd}} = 1$.

Once it exceeds the critical mass $M_{\bullet,\mathrm{crit}}$ in
Eq. \ref{Mcrit}, the black hole enters the second regime, the
`blow-out' phase, which is described by a power-law decline in the
accretion rate.  Fitting the light curves in merger simulations from
\citet{Hopkins06} gives the following parametrization for
$\dot{M}_{\bullet,II}$:
\begin{eqnarray}
\dot{M}_{\bullet,II}(t) = \frac{\dot{M}_{\bullet,{\rm peak}}}{1+(t/t_Q)^{\beta}}
\end{eqnarray}
where $t_Q \propto t_{\mathrm{salp}}$ is the e-folding time,
$\dot{M}_{\bullet,{\rm peak}}$ is the peak accretion rate (given by
$f_{\mathrm{Edd}}$ times the Eddington accretion rate) and $\beta$ is
a parameterized function of the peak accretion rate. In the case that
the initial black hole mass is already larger than the calculated
critical mass, the black hole is not allowed to accrete at the
Eddington rate at all and goes immediately into the blow-out phase. If
the initial black hole is even larger than the calculated final mass,
no quasar phase occurs at all. Note that for calculating the
bolometric luminosity only the accretion rates during the quasar
phases are taken into account (thus ignoring the contribution from the
radio mode accretion):
\begin{eqnarray}\label{Lbol}
L_{\mathrm{bol}} = \frac{\epsilon}{1-\epsilon}\
\dot{M}_{\bullet,\mathrm{QSO}}\ c^2 
\end{eqnarray}
where $\dot{M}_{\bullet,\mathrm{QSO}} = \dot{M}_{\bullet,I}$ in regime
I and $\dot{M}_{\bullet,\mathrm{QSO}} = \dot{M}_{\bullet,I}$ in regime
II.

\subsection{Sub-Eddington limit for the maximum accretion rate}\label{VE}

Observational studies show that the peak in the Eddington ratio
distributions of QSOs is not constant with time; instead, it is found
to be dependent on redshift as well as on black hole mass
(\citealp{Padovani89, Vestergaard03, Shankar04, Kollmeier06, Netzer07,
  Hickox09, Schulze10}).  In particular at low redshifts $z<1$, it has
been claimed that there is a \textit{sub-Eddington} limit for black
hole accretion, which is dependent on black hole mass and redshift
(e.g. \citealp{Netzer07, Steinhardt10}).  The underlying reason for
such a sub-Eddington limit is not obvious but might be related to the
cold gas content of the galaxy. For example, \citet{Hopkins08b} found,
in their semi-empirical models, that allowing `dry' (gas-poor) mergers
to trigger quasar activity would overproduce luminous quasars at low
redshift ($z<1$).  To explore this effect we introduce a limit for the
Eddington ratio at $z \leq 1$, which is dependent on the cold gas
fraction $f_{\mathrm{cold}} =
M_{\mathrm{cold}}/(M_{\mathrm{cold}}+M_{\mathrm{stellar}})$ of the
merged galaxy after the merger. For $f_{\mathrm{cold}} > 0.3$, we
still allow the black hole to accrete up to the Eddington-rate, while
for $f_{\mathrm{cold}} < 0.3$ we assume a simple, linearly decreasing
function for the maximum accretion rate:
\begin{eqnarray}
  f_{\mathrm{Edd}}(f_{\mathrm{cold}}) = 3.3 \times
  f_{\mathrm{cold}} + 0.001
\end{eqnarray}
This lowers the peak accretion rate, and therefore also
correspondingly decreases the accretion rate in the power-law decline
part of the light curve.  Note that the assumption of a limited
accretion rate is also supported by the results of semi-empirical
models by \citet{Shankar11}, which also favor a decreasing
Eddington-ratio with time and a radiative efficiency increasing with
black hole mass.

\subsection{Disk instabilities}\label{DI}

Various observational studies suggest that moderately luminous AGN
are typically not major-merger driven, at $z<1$ (\citealp{Cisternas10,
  Georgakakis09, Pierce07, Grogin05,Salucci99}), and interestingly also at $z
\approx 2$ (\citealp{Kocevski12, Rosario11, Silverman11}) as they do
\textit{not} find more morphological distortions for AGN host galaxies
than for inactive galaxies. This suggests that moderately luminous AGN
may undergo a `main sequence' secular growth, e.g. their nuclear
activity might be additionally driven by disk instabilities. Here we
use a statistic proposed by \citet{Efstathiou82} to quantify disk
stability, based on numerical N-body simulations. They find that the
disk becomes unstable if the ratio of dark matter mass to disk mass
becomes smaller than a critical value, and give the following
parameterization for the onset of disk instabilities:
\begin{eqnarray}   
     M_{\mathrm{disk,crit}}= \frac{v_{\mathrm{max}}^2\ R_{\mathrm{disk}}}{G\
      \epsilon},
\end{eqnarray}
where $M_{\mathrm{disk,crit}}$ is the critical disk mass, above which
the disk is assumed to become unstable, $v_{\mathrm{max}}$ is the
maximum circular velocity, $R_{\mathrm{disk}}$ the exponential disk
length and $\epsilon$ the stability parameter. We use a slightly
smaller value ($\epsilon = 0.75$) than proposed in
\citet{Efstathiou82}, as it results in a better quantitative match
with the observed AGN luminosity function (see section
\ref{Numdens}). However, we find that the number density of AGN is not
very sensitive to the precise value of the stability parameter. A
smaller stability parameter means that the critical disk mass becomes
slightly larger and thus, on average disks become unstable at a later
time, leading to a minor decrease in the number density of
instability-driven AGN. Moreover, we have seen that recent simulations
of isolated disk galaxies tend to indicate an even smaller stability
parameter of $\epsilon \sim 0.6$.
Furthermore, in our model it is assumed that whenever the current disk
mass (consisting of both stars and cold gas) exceeds the critical disk
mass, the bulge component is enlarged by the difference of the
`excess' stellar mass $\Delta M_{\mathrm{disk}} =
M_{\mathrm{disk}}-M_{\mathrm{disk,crit}}$ so that the disk becomes
stable again. We assume that a certain amount of cold gas
(proportional to the excess disk mass) is additionally accreted onto
the black hole triggering an active phase:
\begin{eqnarray}
    M_{\mathrm{disk fuel}} = f_{\mathrm{BH,disk}} \times \Delta M_{\mathrm{disk}}. 
\end{eqnarray}
Here, we adopt $f_{\mathrm{BH,disk}} = 10^{-3}$, motivated by the
local black hole-bulge mass relation.  For the accretion process we
assume a constant Eddington ratio of $f_{\mathrm{edd}}=0.01$ with a
Gaussian distributed scatter of $0.02\ \mathrm{dex}$ (motivated by
observations, D. Alexander private communication). The black hole
accretes as long as there is gas fuel from disk instabilities left
($M_{\mathrm{disk fuel}}>0$) and the accretion rate is calculated by:
\begin{eqnarray}
      \dot{M}_{\bullet,\mathrm{disk}} = 1.26 \times
10^{38} \mathrm{erg/s}\ \frac{1-\epsilon}{\epsilon}
\frac{f_{\mathrm{edd}}}{c^2} M_{\bullet}
\end{eqnarray}
For the total bolometric luminosity, the bolometric luminosities from
the merger driven quasar phase (equation \ref{Lbol}) and from
any disk instablities are summed up:
\begin{eqnarray}
      L_{\mathrm{bol}} = \frac{\epsilon}{1-\epsilon}\
      \left( \dot{M}_{\bullet,\mathrm{QSO}} +
        \dot{M}_{\bullet,\mathrm{disk}} \right)\ c^2.
\end{eqnarray}
In contrast, the studies using the \textsc{Munich} model
(\citealp{Marulli08,Bonoli09}) do not consider disk instabilities for
calculating the AGN bolometric luminosity. In the \textsc{Galform}
model, a similar approach is used to estimate when a disk should
become unstable, but it is then assumed that \emph{all} gas and stars
in the disk are transferred to the bulge component, leading to a much
more dramatic effect. As a result, in their model, disk instabilities
are found to be the \textit{major} physical process responsible for
black hole growth at all redshifts (\citealp{Bower06, Fanidakis10}).

%
 Our simple model is based on simulations of isolated disk
  galaxies, which develop secular internal instabilities. These
  secular instabilities are not expected to be associated with large
  nuclear inflows nor with dramatic morphological or dynamical
  transformation. However, in a cosmological context, the expected
  rapid inflows can lead to more violent ``stream fed'' disk
  instabilities, particularly at high redshift, which may drive BH
  feeding at high rates, and more dramatic morphological/dynamical
  transformation \citep{bournaud:2011,ceverino:2010}.  However, it is
  not known how common such violent instabilities might be in a
  cosmological context, nor how to model their effects within a
  semi-analytic model.  This is an important topic for future work,
  but for the moment we restrict ourselves to the more mild secular
  instabilities. The reader should keep in mind, however, that this
  may represent a minimal prediction for the impact of internal
  instabilities on AGN feeding and spheroid formation.

\subsection{`Heavy' seeding scenario}\label{seed}

The origin of the first massive black holes is still a subject of
intense debate. Currently, there exist two favored seeding mechanisms
(\citealp{Haiman10, Volonteri10}): either black hole seeds could form
out of the remnants of massive Pop III stars
(e.g. \citealp{Madau01,Heger02}) or during the direct core-collapse of
a low-angular momentum gas cloud (e.g. \citealp{Loeb94,
  Koushiappas04,Volonteri08,Volonteri11,Bellovary11}). In the first
case the seeds are expected to have masses of $M_{\mathrm{seed}}
\approx 100-600 M_{\odot}$ (`light' seeding), while in the latter
case, more massive seeds between $M_{\mathrm{seed}} \approx
10^{5}-10^{6} M_{\odot}$ (`heavy' seeding) would be expected. The
detailed physical processes, in particular of the direct
core-collapse, are largely unknown. Unfortunately, observational
constraints in the high-redshift universe are too weak to favor one of
these models. However, future observations of gravitational waves
(LISA, \citealp{Sesana05, Koushiappas06}) or planned X-ray missions
(WFXT: \citealp{Sivakoff10,Gilli10}; IXO), may have the ability to
detect accreting black holes at $z>6$, and thus, will be able to test
these models of the first black holes. Moreover, due to the
exponential growth of the black holes during accretion, it is also
very difficult to use the local population of massive black holes to
recover information about their original masses before the onset of 
accretion. For instance, in the theoretical studies of
\citet{Volonteri08}, and \citet{Volonteri10}, the different seeding
mechanisms are investigated by following the mass assembly using
Monte-Carlo merger trees to the present time. They find that both
models can fit observational constraints at $z=0$ (e.g. the black hole
mass-velocity dispersion relation or the black hole mass function),
when light seeds form already at very early times ($z=20$), or when
heavy seeds evolve later on ($z=5-10$).  Furthermore in a study of
\citet{Tanaka09} they use dark matter halo merger trees,
coupled with a prescription for the halo occupation fraction and they
show that $\approx 100 M_{\odot}$ seed BHs can grow into $10^6
M_{\odot}$ BHs at $z\approx 6$ without super-Eddington accretion, but
only if they form in minihalos at $z \geq 30$.
In our baseline model, seed
black hole masses of $100 M_{\odot}$ were assumed, however, due
to our adopted mass resolution, we obtain very little seeding before
$z \approx 10$. Therefore, we also explore a heavy seeding scenario
with $M_{\bullet,\mathrm{seed}} = 10^5 M_{\odot}$. The different seeding
mechanisms will not affect the $z=0$ black hole mass and AGN
population, as initial seed masses are compensated by gas accretion
growth processes by orders of magnitude. Only the black hole
distribution and QSO luminosity function at high redshifts will be
influenced by this modification. Furthermore, some studies have
suggested that low mass dark matter halos may not be able to produce
massive seeds (\citealp{Menci08, Volonteri11_2}), as the potential
well might be too weak for collapse to occur. Therefore in the `heavy
seeding' model we additionally adopt a halo mass limit of $2 \times
10^{11} M_{\odot}$, below which no black hole seeds are inserted.

Furthermore, many observations (e.g. \citealp{Walter04, Peng06,
  McLure06, Schramm08}) suggest that the black hole-to-bulge mass
ratio was larger at higher redshifts than expected from the local
black hole-bulge mass relation. This eventually implies that black
holes were accreting more gas and thus, growing faster than the
corresponding bulges at high redshifts than at lower ones. Therefore,
besides assuming an evolving black hole-bulge mass relation (see the
z-dependent $\Gamma$ parameter in eq. \ref{mbhfinal}) we additionally
adopt a larger scatter $\sigma_{\bullet,\mathrm{accr}}$ for the
accreted mass onto the black holes at high redshifts. This means that,
when calculating the final black hole mass
$M_{\bullet,\mathrm{final}}$, a larger Gaussian distributed scatter is
applied with a value of $\sigma_{\bullet,\mathrm{accr}}=0.6$ for
$z>4$. At redshift $z<4$ the original scatter value
$\sigma_{\bullet,\mathrm{accr}}=0.3$ is applied. We find that we still
recover a tight relationship between BH mass and bulge mass at $z=0$,
in agreement with observations. This is in agreement with the results
of \citet{Hirschmann10}, who showed that a large scatter in black hole
mass at fixed bulge mass ($\sigma = 0.6\ \mathrm{dex}$) at high
redshift ($z=3$) will decrease towards the observed present-day value
due to mergers.

\subsection{Summary of Model Variants}

In the course of this study, we investigate the effects of the
outlined modifications on the AGN/black hole evolution one-by-one and
in various combinations. We consider the following six different
models:
\begin{enumerate}[1.]
\item{{\bf{FID}}: \textbf{Fid}ucial, standard accretion model (\ref{FID})}
\item{{\bf{VE}}: \textbf{V}arying sub-\textbf{E}ddington limit for the
    maximum accretion rate $f_{\mathrm{edd}}$ (\ref{FID} \& \ref{VE})}
 \item{{\bf{DI}}: Additional accretion due to \textbf{D}isk
     \textbf{I}nstabilities (\ref{FID} \& \ref{DI})}
\item{{\bf{SH}}: Heavy \textbf{S}eeding mechanism with a \textbf{H}alo
    mass limit (\ref{FID} \& \ref{seed})}
\item{{\bf{DISH}}: 
    \textbf{D}isk \textbf{I}nstabilities \& Heavy \textbf{S}eeding mechanism with a \textbf{H}alo
    mass limit (\ref{FID},\ref{DI} \& \ref{seed})}
\item{{\bf{VEDISH}}: Best-fit model including a \textbf{V}arying sub-\textbf{E}ddington limit,
    \textbf{D}isk \textbf{I}nstabilities \& a heavy \textbf{S}eeding
    mechanism with a \textbf{H}alo mass limit
    (\ref{FID}, \ref{VE}, \ref{DI} \& \ref{seed})} 
\end{enumerate}

\section{Properties of nearby galaxies and black holes}\label{obs} 

\begin{figure*}
\begin{center}
  \epsfig{file=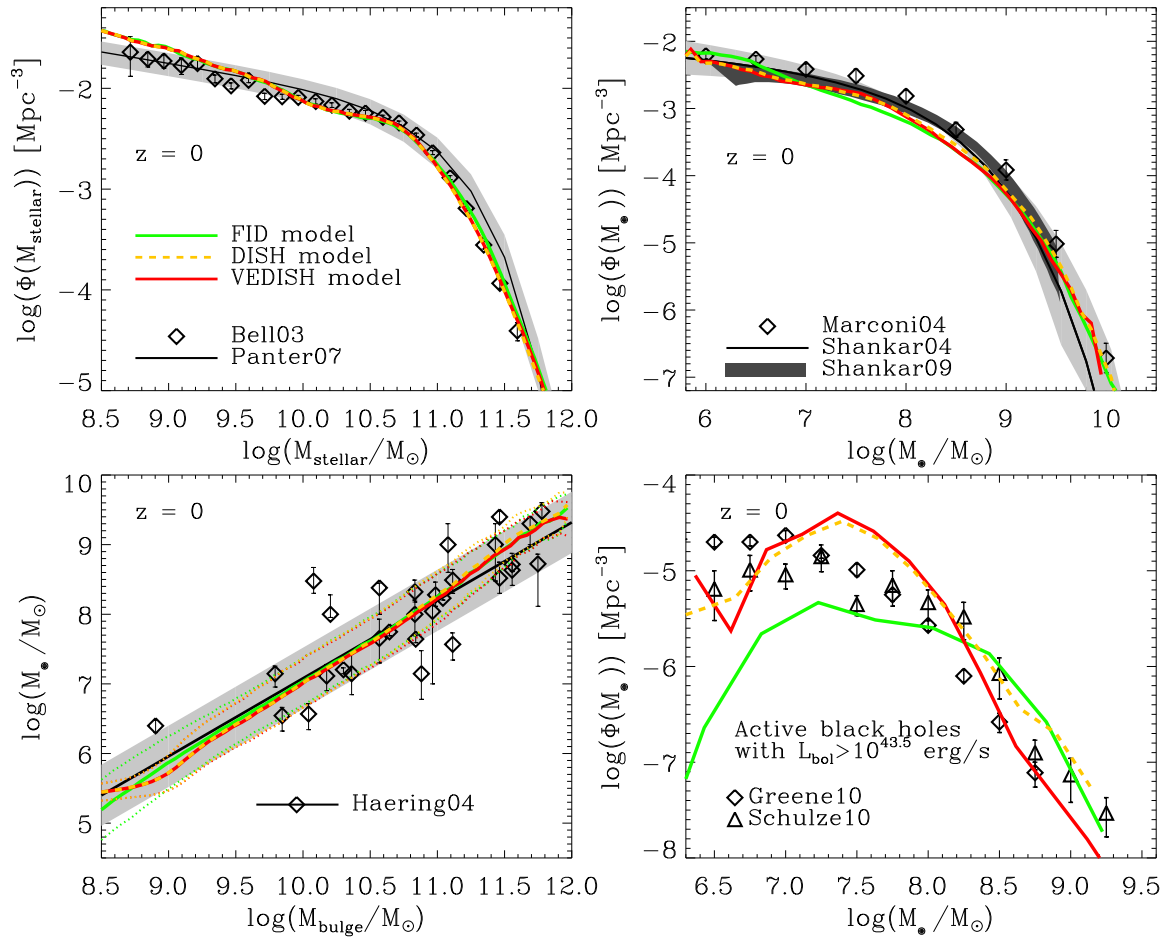, width=0.9\textwidth}
  \caption{Comparison of SAM results to observations for black hole
    and galaxy properties of the local Universe. The
    solid lines are the FID (green) and the VEDISH (red) model, the
    dashed, orange lines the DISH model.  Upper left panel: the
    stellar mass function. The black solid line and the black symbols
    show observational results from \citet{Panter07} and \citet{Bell03},
    respectively. Upper right panel: the black hole mass
    function. Observational estimates \citep{Marconi04, Shankar04,
      Shankar09} are illustrated by the black solid lines, the
    grey shaded areas and the black symbols. Bottom left panel: black
    hole-bulge mass relation. The black solid line and symbols
    correspond to the  
  observed relation by \citet{Haering04} and the grey
  shaded area show the 1-$\sigma$ scatter of the observational
  data. Bottom right panel: the mass function of active black 
  holes with bolometric luminosities above $L_{\mathrm{bol}}<10^{43.5}\
  \mathrm{erg/s}$. Open symbols show observations from
  \citet{Greene09} and \citet{Schulze10}. Disk instabilities increase
  the number of active low mass black holes and the gas-dependent
  Eddington-limit decreases the number of active high mass black holes.}
 {\label{Propz0}}
\end{center}
\end{figure*}

In Fig. \ref{Propz0} we compare different galaxy and black hole
properties from the FID, the DISH and the VEDISH model to observations
of the local Universe. We do not explicitly show the predictions of
the VE, the DI and the SH model separately as they do not result in a
stronger deviation from the FID model than the DISH or the VEDISH
model.

The upper left panel in Fig. \ref{Propz0} shows the modeled stellar
mass functions compared to observations from \citet{Bell03} and
  \citet{Panter07}. The modifications in the DISH and the VEDISH model hardly
show any variation from the FID model and thus, for stellar masses
larger than $10^9 M_{\odot}$ we obtain a reasonably good match to the
observational data. However, low-mass galaxies ($<10^9 M_{\odot}$) are
slightly over-predicted, a common feature of most (all) current
semi-analytic models (e.g. \citealp{Bower06,DeLucia07}) and still a
subject of on-going research
(e.g. \citealp{Guo11,Bower11,Wang11,Menci12}). Stronger supernova
feedback, a modified star formation law or a different cosmological
model are currently considered as possible solutions for this problem.

The present-day black hole mass function is depicted in the upper
right panel of Fig. \ref{Propz0}. Our model predictions are compared
with observational estimates from \citet{Shankar04},
\citet{Marconi04} and \citet{Shankar09} (see the review of
\citealp{Shankar09_rev} for more details). We find reasonably good
agreement of the SAM predictions with the observations for the whole
black hole mass range. Deviations between the different SAM models are
negligible, indicating that neither the growth of black holes by disk
instabilities nor the limited accretion rates at low redshifts 
influence the black hole mass function significantly. In contrast to
our result, in many SAM studies an excess of very massive black holes
can be seen (e.g. \citealp{Fontanot06, Malbon07, Marulli08,
  Fanidakis10}), which might be caused by too-efficient Radio-mode
accretion as discussed by \citet{Fontanot11}.  

\begin{figure*}
\begin{center}
  \epsfig{file=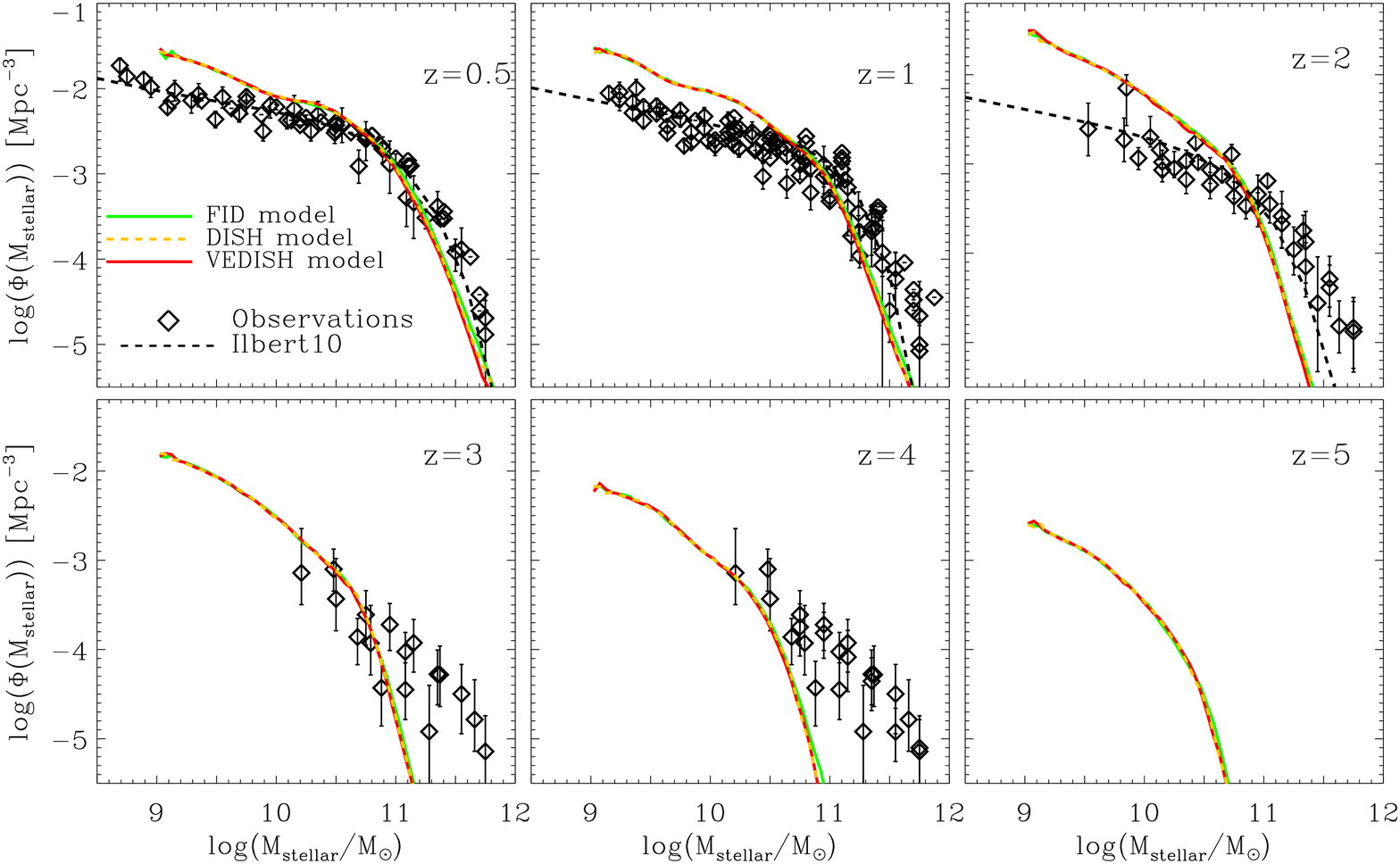, width=0.9\textwidth}
  \caption{Redshift evolution of the stellar mass function
    ($z=0.5,1,2,3,4,5$). The solid lines show the FID (green) and the
    VEDISH (red) model and the orange dashed lines the DISH model. The
    black dashed lines depict observations from \citet{Ilbert10} and
    the symbols correspond to observations from a set of observational
    studies (\citet{Perez08, Bundy05, Drory04, Fontana06,
      Marchesini08}). As in most SAMs, the number of low mass
      galaxies is overestimated (at all redshifts), while the number
      of high mass galaxies is underestimated at high z. }
          {\label{Evol_stellmass}}
\end{center}
\end{figure*}
 
The present-day relation between black hole and bulge mass is shown in
the lower left panel of Fig. \ref{Propz0}, with our model predictions
compared with the observational relation derived by \citet{Haering04}.
All of the model variants reproduce a tight relationship between black
hole mass and bulge mass, which results from the self-regulated BH
growth assumed in our model. The slight upturn at low bulge masses in
the DISH and the VEDISH models is due to the heavy seeding mechanism,
where by construction no black hole masses below $10^5 M_{\odot}$ can
exist.

Finally, the lower right panel of Fig. \ref{Propz0} shows the active
black hole mass function predicted by our model, compared with
observational data from \citet{Greene09} and \citet{Schulze10}. Note
that --- consistent with these observational studies --- we define
``active'' BH here as having a bolometric luminosity greater than
$10^{43.5}\ \mathrm{erg/s}$.  For $M_{\bullet} < 10^{8.3}M_{\odot}$,
the accretion due to disk instabilities (DISH/VEDISH models) increases
the fraction of active black holes by almost one order of magnitude
compared to the FID model. In contrast, for $M_{\bullet} >
10^{8.3}M_{\odot}$ the limited gas accretion in the VEDISH model
reduces the active fraction of massive black holes compared to the
DISH and the FID model. Overall, the VEDISH model provides a better
match to the observational data than the FID or the DISH models.

\begin{figure}
\begin{center}
  \epsfig{file=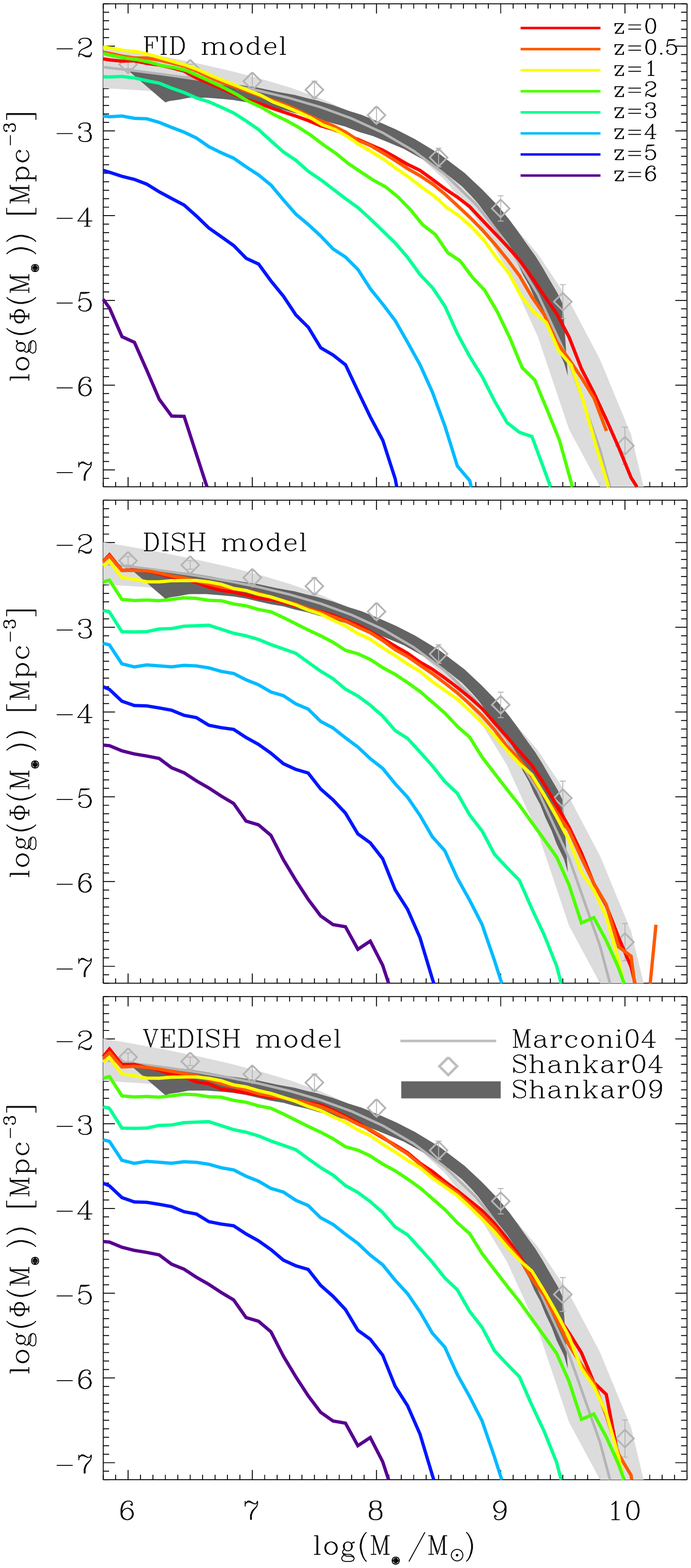, width=0.4\textwidth}
  \caption{Redshift evolution of the black hole mass function. The
    colored lines illustrate the SAM results of the FID (upper panel),
    DISH (middle panel) and the VEDISH (low panel) model at different
    redshifts. The grey lines and symbols correspond to observational
    estimates \citep{Marconi04,Shankar04,Shankar09} at
    $z=0$. Massive black hole seeds significantly increase the number
    of black holes with masses between
    $10^6<M_{\bullet}<10^8M_{\odot}$ at $z>4$.}  {\label{Evol_mbh}}
\end{center}
\end{figure}

\section{Galaxy and black hole properties at higher redshift}\label{BHpropevol}

Fig. \ref{Evol_stellmass} illustrates the stellar mass functions at
different redshifts ($z=0.5,1,2,3,4,5$) as indicated in the legend of
each panel. We compare our SAM predictions with the results of
different observational studies (\citealp{Perez08, Bundy05, Drory04,
  Fontana06, Marchesini08,Ilbert10}).  At all redshifts, there is no
significant difference between the different model variants. However,
compared to observations, the high mass end is under-predicted, while
the low-mass end is over-predicted by the SAMs. This discrepancy is
again a well-known problem \citep{Fontanot09,Marchesini07}.
\citet{Fontanot09} showed this for the \textsc{Morgana},
\textsc{Munich} and the S08 model, and \citet{Guo11} found the same
for the latest version of the \textsc{Munich} model. The discrepancy
at the high-mass end may be related to systematic errors or scatter in
the photometric stellar mass estimates from observations ---
\citet{Fontanot09} showed that when stellar mass errors of $0.25$ dex
were convolved with the model results, the SAMs agreed reasonably well
with the available stellar mass function compilations at least to
$z\sim 3$. Moreover, \citet{Somerville11} showed that the SAM
presented here agrees with the observed rest-frame $K$-band luminosity
function at the bright end up to $z\sim 3$.  The excess of low mass
galaxies at high redshift is also seen in numerical cosmological
hydrodynamic simulations \citep{Dave11}, showing that this problem is
not peculiar to SAMs. Instead, it may be an indication that the star
formation or supernova feedback recipes that are commonly adopted in
both SAMs and numerical simulations require revision (Caviglia \&
Somerville, in prep.), or that the underlying cosmological model
differs from the Cold Dark Matter paradigm (e.g., warm dark matter;
\citealp{Menci12}).

\begin{figure}
\begin{center}
  \epsfig{file=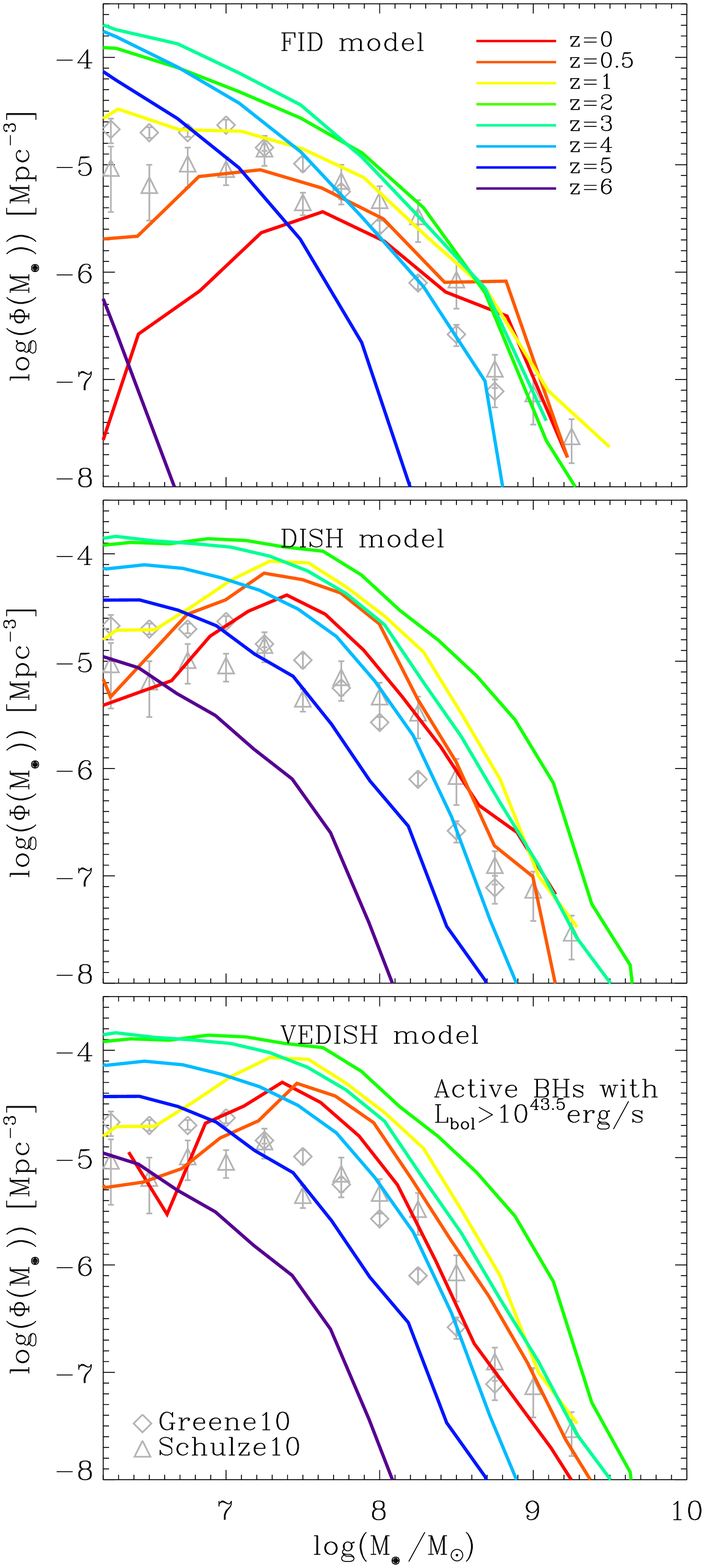, width=0.4\textwidth}
  \caption{Redshift evolution of the black hole mass function of active
    black holes with bolometric luminosities larger than
    $L_{\mathrm{bol}}>10^{43.5}\ \mathrm{erg/s}$. Colored
  lines illustrate the SAM results of the FID (upper panel), the DISH
  (middle panel) and the VEDISH (lower panel) model at different
  redshifts. For comparison, the grey
  symbols illustrate observations of the active black hole mass
  function at $z=0$ \citep{Greene09, Schulze10}. Massive seeds
  increase the number of active black holes at high redshift and disk
  instabilities increase the number of active low mass black holes at
  low redshift.} 
 {\label{Evol_activembh}}
\end{center}
\end{figure}

Fig. \ref{Evol_mbh} shows the black hole mass function at different
redshift steps ($z = 0, 0.5, 1, 2, 3, 4, 5, 6$) for the FID, DISH, and
VEDISH models. We show the observational estimates at $z=0$ to guide
the eye. At redshifts $z < 3$, the different model assumptions in the
DISH and the VEDISH models do not influence the evolution of the black
holes. However, turning to higher redshift $z \geq 3$, the main
difference between the FID model and the DISH/VEDISH models is the
larger number density of black holes more massive than $M_{\bullet} >
10^6 M_{\odot}$. This can be explained by the `heavy' seeding scenario
and the large scatter in the accreted gas mass onto the black hole at
$z>4$. This leads to larger black hole masses and faster growth at
early redshift than in the FID model. Towards lower redshift, however,
this effect dissappears as the subsequent growth by gas accretion
overcomes the seed black hole masses by orders of magnitude.  This
trend is even more pronounced in the model of \citet{Fanidakis10}, as
they allow the black holes to accrete at super-Eddington rates:
e.g. at redshift $z=6$, black holes with $M_{\bullet} = 10^6
M_{\odot}$ have a number density of $\log \Phi =
-2.7\ \mathrm{Mpc}^{-3}\ \mathrm{dex}^{-1}$, whereas we obtain a
number density of only $\log \Phi =
-4\ \mathrm{Mpc}^{-3}\ \mathrm{dex}^{-1}$ in the VEDISH model.

The evolution of the active fraction of black holes is shown in
Fig. \ref{Evol_activembh}. The three panels correspond to the
different models (FID: upper panel, DISH: middel panel and VEDISH:
lower panel), where colored lines illustrate the model results at
different redshifts ($z=0,0.5,1,2,3,4,5,6$). We consider only AGN with
bolometric luminosity larger than $10^{43.5}\ \mathrm{erg/s}$. The
grey symbols show the local observations. We find that the evolution
of the active black hole fraction varies from model to model, implying
that our modifications to the recipes for BH formation and evolution
have a significant influence on the active fraction at all
redshifts. Comparing the DISH model with the FID model we can see two
effects: At high redshifts $z \geq 3$, the number of active black
holes with masses between $10^6< M_{\bullet} < 10^8M_{\odot}$ is
greatly increased due to the heavy seeding mechanism and the large
scatter in the accreted gas mass. At low redshifts $z \leq 1$, the
number of active black holes with masses below $10^{8.3}M_{\odot}$
rises as a consequence of the additional gas accretion due to disk
instabilities. Furthermore, as already seen in Fig. \ref{Propz0}, the
limited accretion rate in the VEDISH model reduces the fraction of AGN
with massive black holes ($>10^{8.3} M_{\odot}$) at $z<0.5$.

\section{Number density evolution of AGN}\label{Numdens} 

\begin{figure*}
\begin{center}
  \epsfig{file=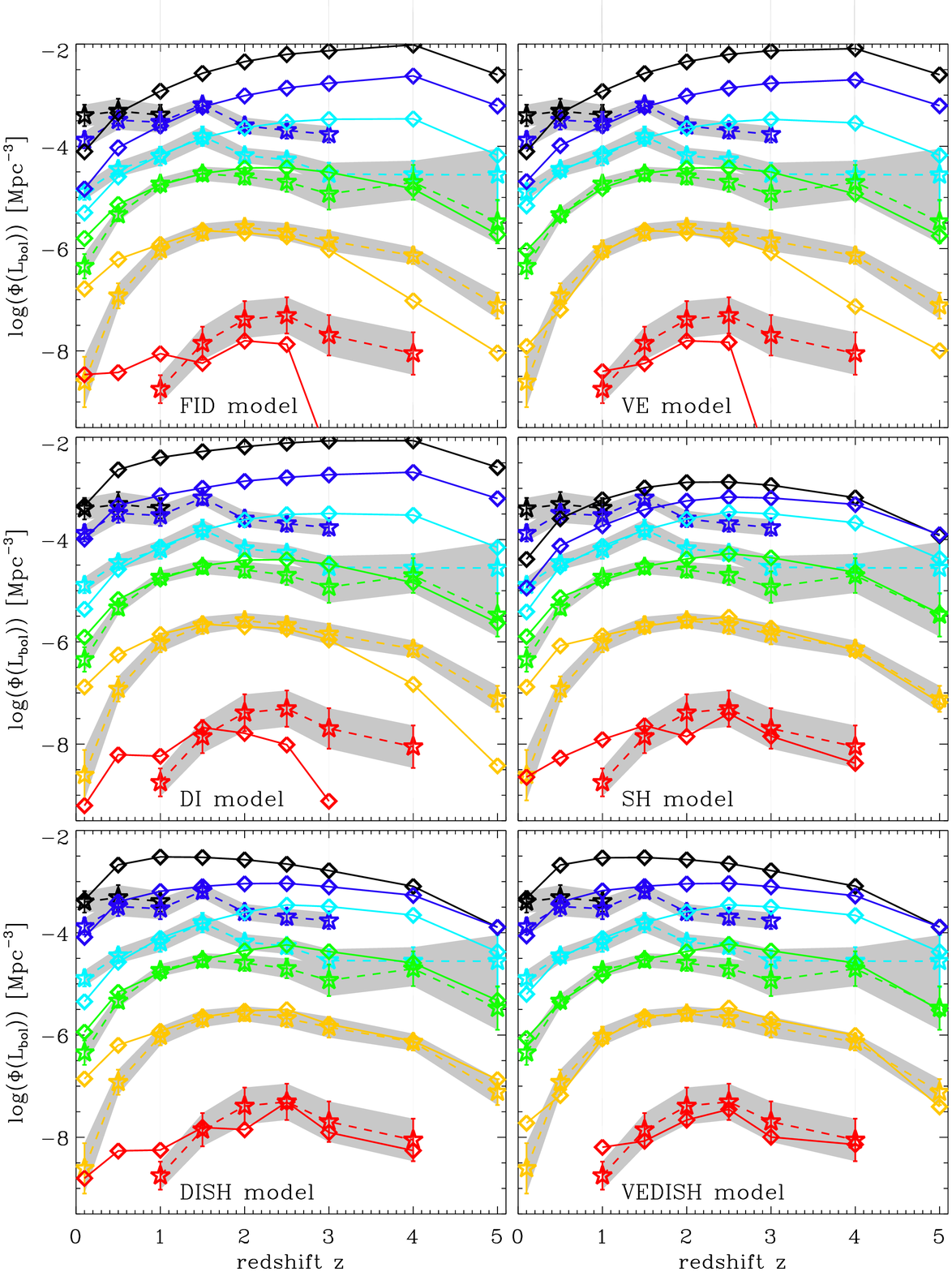, width=0.9\textwidth}
  \caption{Number densities of AGN versus redshift for the six
    different SAM models (FID, VE, DI, SH, DISH and VEDISH) as
    indicated in the legend. Different colors illustrate different
    bolometric luminosity bins: red: $47.5<\log(L_{\mathrm{bol}})$,
    yellow: $46.5<\log(L_{\mathrm{bol}})<47.5$, green:
    $45.5<\log(L_{\mathrm{bol}})<46.5$, light blue:
    $44.5<\log(L_{\mathrm{bol}})<45.5$, dark blue:
    $43.5<\log(L_{\mathrm{bol}})<44.5$, black:
    $42.5<\log(L_{\mathrm{bol}})<43.5$. Solid lines and open squares
    show the corresponding model predictions; dashed lines and stars
    together with grey shaded areas indicate the observational
    compilation from \citet{Hopkins07}. While the FID model shows the
    opposite of ``downsizing'' behavior, with luminous objects forming
    late and low-luminosity objects forming early, we obtain a fairly
    good match to the observations for the VEDISH model.}
          {\label{Numdens_all}}
\end{center}
\end{figure*}

The different panels in Fig. \ref{Numdens_all} show the redshift
evolution of the AGN number densities as a function of the bolometric
luminosity, for the six different models.
In this section, our SAM predictions are compared to the observational
compilation from \citet{Hopkins07}. In their study, they convert the
AGN luminosities from different observational data sets and thus, from
different wavebands (emission lines, NIR, optical, soft and hard
X-ray) into bolometric ones.  They assume a luminosity dependence of
the obscured fraction (the less luminous the more obscured) and the
same number of Compton-thick ($N_H > 10^{24}\ \mathrm{cm}^{-2}$) and
Compton-thin ($10^{23}\ \mathrm{cm}^{-2}< N_H <
10^{24}\ \mathrm{cm}^{-2}$) AGN. However, there are many aspects of
the obscuration corrections that are still being vigorously
debated. Some recent studies suggest that the obscured fraction is
dependent on both luminosity and redshift
(\citealp{Hasinger08,Fiore12}), in contrast with the non-redshift
dependent model of \citet{Hopkins07}. There are also uncertainties
surrounding the dust correction for the UV luminosity;
\citet{Hopkins07} compute the amount of dust (and therefore
extinction), by adopting an $N_H$ distribution from X-ray
observations, and a Galactic dust-to-gas ratio. However, it has been
shown that AGN absorbers do not have a Galactic dust to gas ratio
(\citealp{Maiolino01, Maiolino04}). The result is that they probably
over-estimate the extinction, which might result in slightly higher
luminosities for the optically selected quasars (F. Fiore, personal
communication). Because of these uncertainties, we both compare the
obscuration-corrected observational compilation of \citet{Hopkins07}
with our unobscured model predictions, and in Section~\ref{Xlum} we
attempt to correct our model predictions for obscuration and compare
with recent soft and hard X-ray measurements of the AGN luminosity
function.


The upper left panel in Fig. \ref{Numdens_all} shows the FID
model. The number densities at the peak of each luminosity bin are in
quantitative agreement with the observations implying that the FID
model reproduces the correct order of magnitude of AGN number
densities in the different luminosity bins. However, the observed time
evolution of the peaks of the different luminosity classes are not
correctly predicted by the FID model, which shows the typical
``hierarchical'' behavior in which the number of low-luminosity
objects peaks earlier than the number density of higher luminosity
objects. We can summarize this problem more quantitatively as follows:
\begin{itemize}
\item{$z < 2$: over-prediction of AGN
     with $\log(L_{\mathrm{bol}}) > 46$}\label{lumzlow}
\item{$z < 2$: under-prediction of AGN with
    $\log(L_{\mathrm{bol}}) < 46$}\label{lesslumzlow}
\item{$z > 3$: under-prediction of AGN 
    with $\log(L_{\mathrm{bol}}) > 46$} \label{lumzhigh}
\item{$z > 3$: over-prediction of AGN with
    $\log(L_{\mathrm{bol}}) < 45$} \label{lesslumzhigh}
\end{itemize}

As the fiducial model reproduces the black hole mass function at z=0,
we can assume that there are the correct number of black
holes. Therefore the first point suggests that at low redshifts either
too high a fraction of massive black hole are accreting or these
massive black holes are accreting at rates that are too high.
Moreover, assuming that activity is triggered by merger events implies
that the natural decrease in the major merger rate is not sufficient
to produce the observed steep decline in the AGN number densities. It
has been shown that the galaxy merger rate predicted by the S08 models
matches observations (\citealp{Jogee08}, \citealp{Lotz11}) so this is
not likely to be the cause of the discrepancy.  
The low number densities of moderately luminous AGN may indicate,
however, that the AGN activity might not only be triggered by merger
events, but also by secular evolution processes. The deviations at
high redshift may be a consequence of massive black holes forming too
late in the SAM as well as possibly the non-redshift-dependent dust
obscuration correction.  The excess of moderate and low luminosity AGN
at high redshift may be partly due to the similar over-prediction of
low-mass galaxies in the SAM, which we already discussed in
Section~\ref{BHpropevol}.
\begin{figure*}
\begin{center}
  \epsfig{file=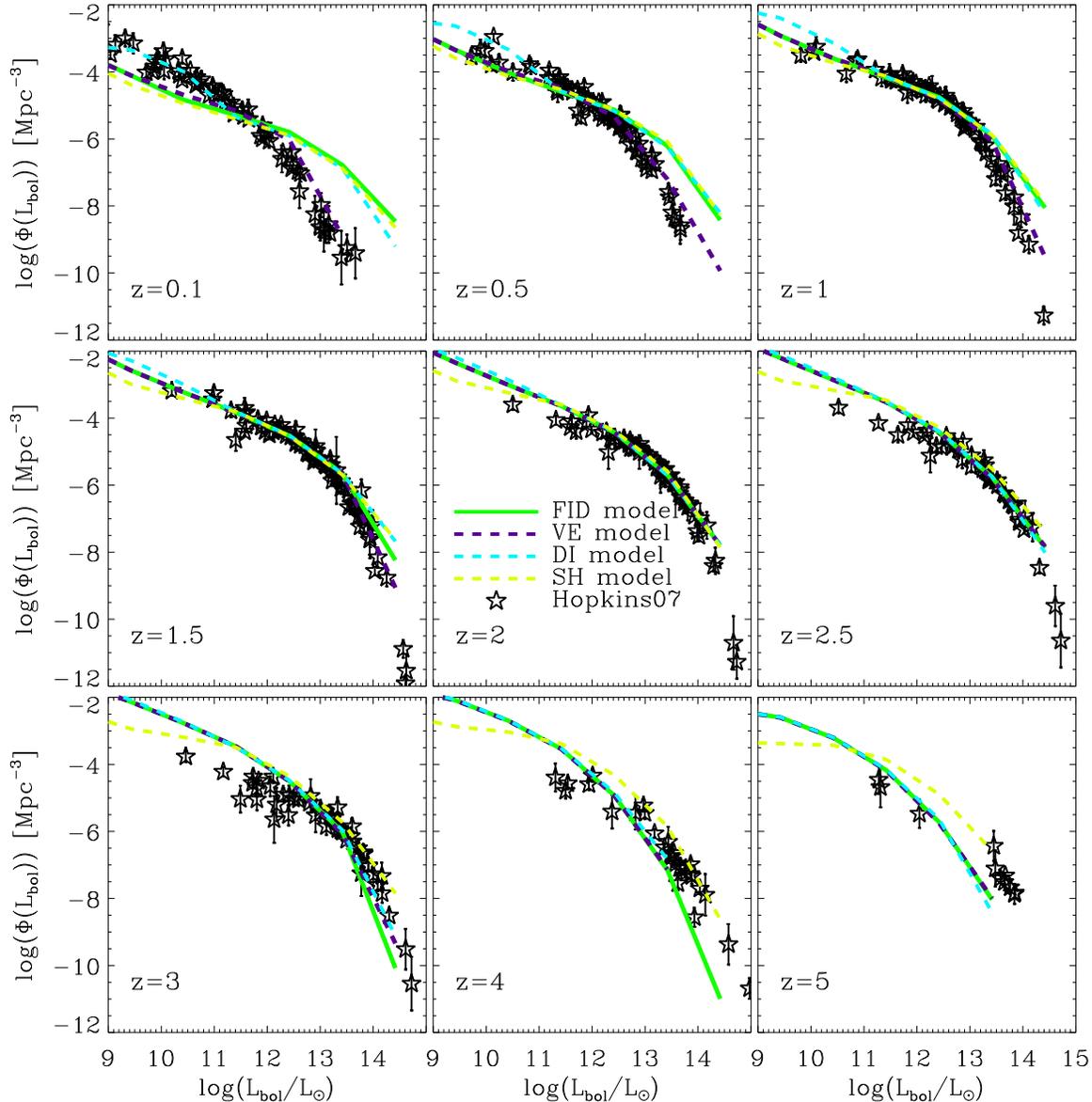,
    width=0.9\textwidth} 
  \caption{Bolometric quasar luminosity functions at different
    redshifts. Black stars show the observational compilation of
    \citet{Hopkins07}. The green solid lines correspond to the output
    of the FID model, the purple, dashed lines show the VE model, the
    light blue, dashed lines the DI model and the dark blue, dashed
    lines the SH model.}  {\label{QLF2}}
\end{center}
\end{figure*}
\begin{figure*}
\begin{center}
  \epsfig{file=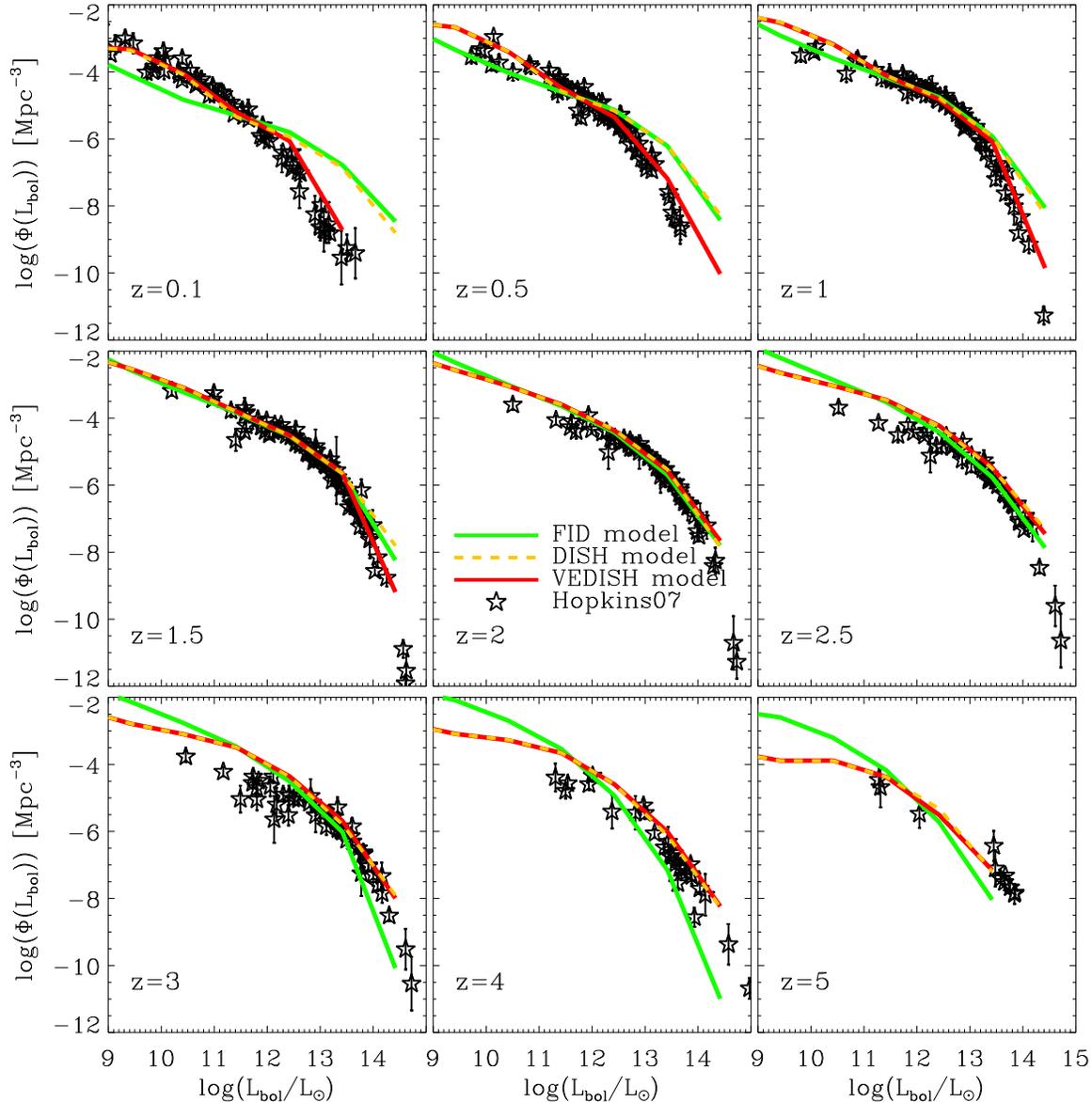,
    width=0.9\textwidth} 
  \caption{The same as in Fig. \ref{QLF2}, but for different SAM models: the
    green solid lines correspond to the output of the FID model, the
    orange, dashed lines show the DISH model and the red, solid lines
    illustrate the results from the VEDISH model. For the VEDISH
    model, a reasonably good agreement with observations is obtained.}
 {\label{QLF}}
\end{center}
\end{figure*}

We experimented with whether we could achieve better reproduction of
the observed downsizing of BH activity by just modifying the values of
some free parameters in the FID SAM, without changing any physical
ingredients. For example, we tested the effect of varying the strength
of Supernova and radio-mode feedback (as suggested in
\citealp{Fontanot06}). We find that for stronger supernova feedback
(doubling the normalization of the mass loading factor for SN-driven
winds $\epsilon_{\mathrm{SN}}^{0}$) we achieve a decrease in the
number density of AGN at \textit{all} redshifts, resulting in overall
a worse match to the observations. Even if the number density of
moderately luminous objects ($L_{\mathrm{bol}} <
10^{45}\ \mathrm{erg/s}$) at high redshifts decreases (towards the
observational data), this is not sufficient to achieve a reasonably
good match to observations in this range. Increasing the strength of
the radio-mode feedback reduces the number of luminous AGN
($L_{\mathrm{bol}} > 10^{45}\ \mathrm{erg/s}$) between redshifts
$0<z<4$, again resulting in worse agreement with the observational
compilation than for the FID model itself. If we assume `halo
quenching' instead of the radio-mode feedback model, i.e. no cooling
is allowed above a certain threshold halo mass
($M_{\mathrm{halo,thres}} = 10^{12} M_{\odot}$), we obtain a decrease
of the number of luminous AGN, but again at \textit{all} redshifts,
and not redshift dependent as observed. To increase the number density
of moderately luminous objects at low redshifts, we varied the
timescale ($t_Q$) in the power-law decline phase of a quasar
episode. A study by \citet{Marulli08} has shown that a power-law
decline growth phase in their quasar mode \textit{does} increase the
number density of moderately luminous AGN at low redshift, resulting
in a better agreement with the observations. However, within our study
it is found that varying $t_Q$ is not sufficient to match the
observations. Therefore, we can conclude that downsizing cannot be
reproduced solely by varying the free parameter values of our FID
model. Instead, we now present the influence of the additional
modifications for black hole formation and growth as outlined in
Section \ref{model}.


The result for the VE model (i.e. assuming a sub-Eddington accretion
limit dependent on the cold gas fraction) is illustrated in the upper
right panel of Fig. \ref{Numdens_all}. For bolometric luminosities
larger than $10^{46}\ \mathrm{erg/s}$, the cold gas fraction dependent
sub-Eddington limit clearly reduces the number density of AGN. We find
that we are able to reproduce the observed steep decline in the number
densities of bright AGN at $z<2$ reasonably well in this way. This
supports the idea that the cold gas content of a galaxy may regulate
the efficiency of black hole accretion, in particular the maximum
accretion rate that can be reached in a merger. We may speculate that
low cold gas densities lead to smaller viscosities so that it takes
longer for the gas to lose its angular momentum and thus, to be
accreted onto the black hole. Our results imply that even when massive
black holes experience major mergers, if the gas fractions in their
host galaxies are low they may not produce luminous AGN because they
never accrete at close to the Eddington rate.

The middle left panel of Fig. \ref{Numdens_all} shows the results of
the DI model, assuming an additional BH accretion mode due to disk
instabilities. For a luminosity range of $43 < \log(L_{\mathrm{bol}})
< 45$, the number of AGN is increased, resulting in a reasonably good
match with the observational compilation for $z<1.5$. For the lowest
luminosity bin, however, the number of AGN is now
over-predicted. Nevertheless, this additional accretion mechanism
seems to play an important role for triggering the activity of faint
AGN (consistent with observational results e.g. \citealp{Salucci00}). In strong
contrast to our model, black hole accretion due to
disk instabilities in the \textsc{Galform}-model
(e.g. \citealp{Fanidakis10}) provides the \textit{major} contribution
to AGN number densities for \textit{all} luminosities and at
\textit{all} redshifts.
In the \textsc{Munich}-models as presented by \citet{Marulli08} and
\citet{Bonoli09}, black hole accretion due to disk instabilities is
not accounted for at all, but they still slightly under-predict the
number of moderately luminous AGN, even in their best-fit model.  This
further supports the need for BH accretion driven by secular evolution
processes, in addition to mergers.


The effect of a heavy seeding mechanism together with a halo mass
limit for black hole formation (SH model) is illustrated in the middle
right panel of Fig. \ref{Numdens_all}. The number of bright AGN at
high redshift is increased and can match the observational data. This
is a consequence of large seed black hole masses and a large scatter
in the accreted gas mass. As we cap BH growth at the Eddington rate,
having more massive seeds means that these early black holes can grow
faster, leading to a larger number of massive and active black holes
at early times than in the FID, VE and DI models. Our result indicates
that black holes probably have to undergo a phase of very rapid growth
at high redshifts ($z>5$), even if it is still unknown whether and how
such massive seed black holes can form out of direct core-collapse or
whether less massive seeds have to accrete at super-Eddington
rates. However, we find that assuming even more massive black hole
seeds with masses of $M_{\bullet,\mathrm{seed}} = 10^6 M_{\odot}$
results too few moderately luminous AGN at high redshift ($z \approx
5$) in our model.
The halo mass limit for seed black hole formation also reduces the
number density of faint AGN, as black holes in low-mass halos are not
allowed to form and to accrete gas. However, even if this second
effect results in better agreement with the observational data, it
does not seem to be fully sufficient for reproducing them. This might
indicate that the dependence of seed mass on halo mass is more complex
than we have assumed in this simple model, or might be be due to
redshift-dependent obscuration, which has not been accounted for in
the observational comparison that we are using here (see Section
\ref{Xlum}).

Finally, the combination of the individual modifications which have
been discussed so far is presented by the DISH and the VEDISH model
(lower left and lower right panel of Fig. \ref{Numdens_all}). We find
that the changes in the AGN number densities seen for the individual
modifications sum in a straightforward way, without significantly
influencing each other. Thus, the VEDISH (=``best-fit'') model is able
to reproduce the observed downsizing trend fairly well, and can
predict the \textit{correct time-evolution} of the peaks of the
luminosity dependent number density curves. However, even in our
best-fit model, the number of faint AGN is still over-predicted at
redshifts between $4>z>2$. This might be due to the difficulty of
detecting these objects at high redshifts, either because they may not
be easily recognizable as AGN, or because of obscuration.

\section{The AGN luminosity function}\label{AGNlum} 

\subsection{Bolometric luminosities}\label{bollum}

\begin{figure*}
\begin{center}
  \epsfig{file=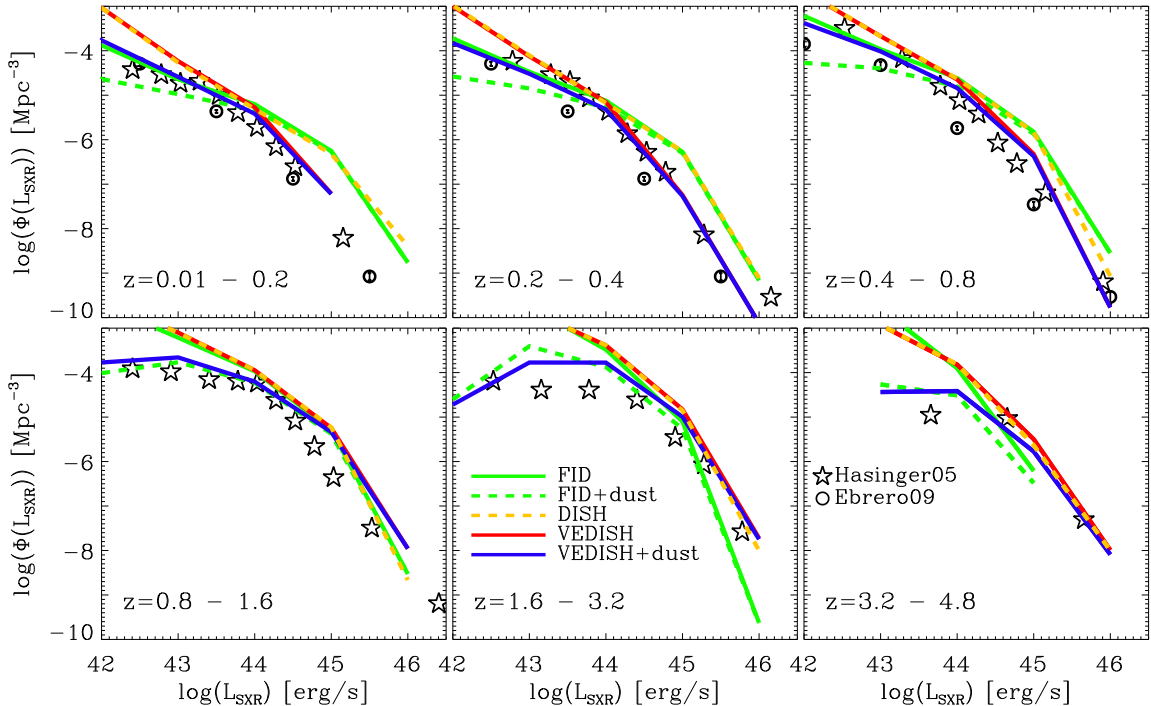,
    width=0.9\textwidth} 
  \caption{Soft X-ray AGN luminosity function for different redshift
    ranges. Black stars and circles show the observational data from
    \citet{Hasinger05} and \citet{Ebrero09}, respectively. The solid
    green, dashed orange and solid red lines correspond to the results
    of the FID, the DISH and the VEDISH model and the blue solid lines
    illustrate the results of the VEDISH model with an obscuration
    correction according to \citet{Hasinger08}. The VEDISH model
    including the obscuration correction is in fair agreement with the
    observational data.}  {\label{SXR}}
\end{center}
\end{figure*}
\begin{figure*}
\begin{center}
  \epsfig{file=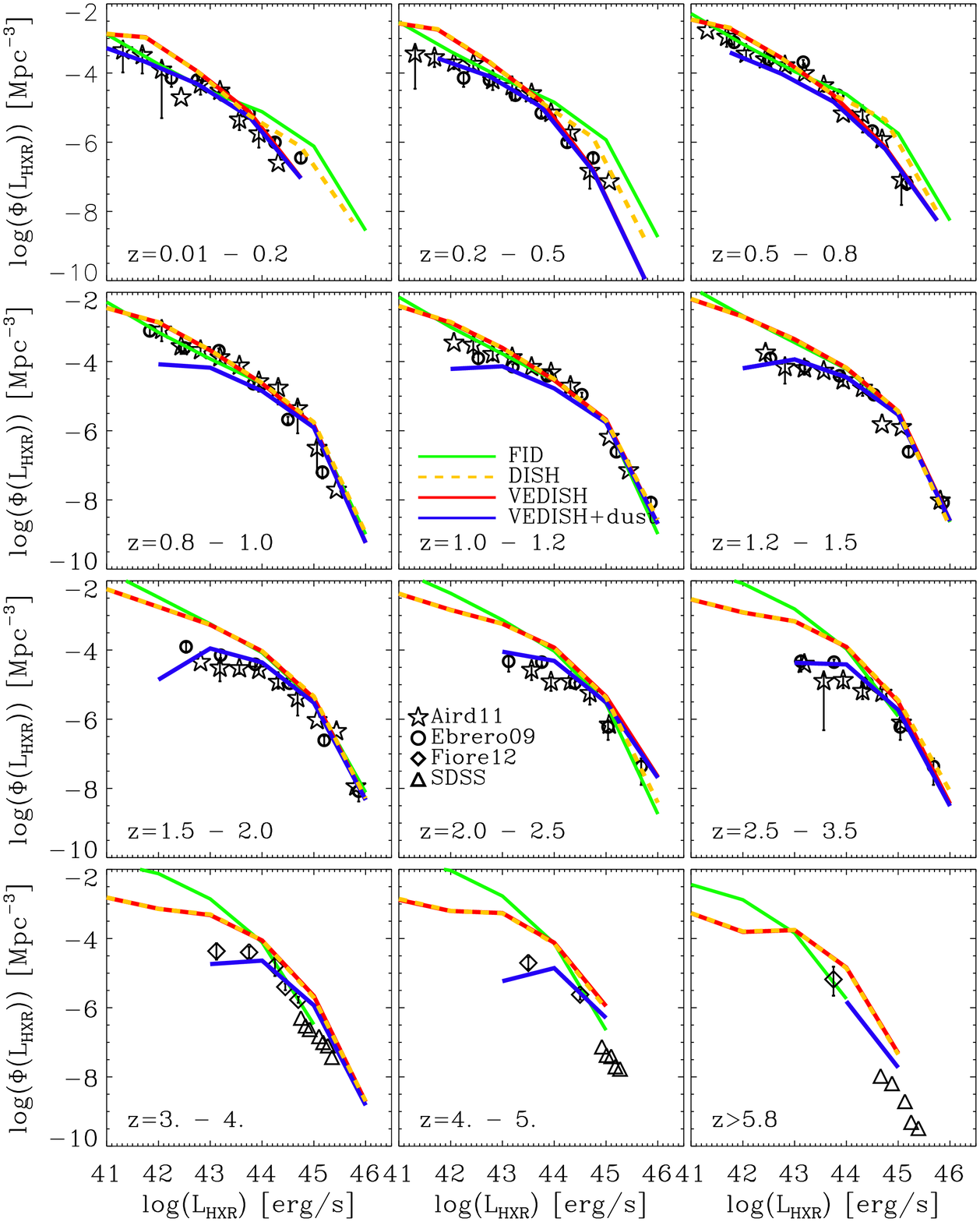,
    width=0.9\textwidth} 
  \caption{Hard X-ray AGN luminosity function for different
    redshift ranges. Black symbols show
    the observational data of \citet{Ebrero09,Aird10,Fiore12} and data from
    the SDSS (optical luminosity is converted into X-ray luminosity as
    in \citet{Fiore12}). 
    The solid green, dashed orange and solid red lines
    illustrate the output of the FID, the DISH and the VEDISH
    model. The blue solid lines illustrate the results of the VEDISH
    model with an obscuration correction according to \citet{Hasinger08}. The
    latter is able to predict the observed hard X-ray LF reasonably
    well. } 
 {\label{HXR}}
\end{center}
\end{figure*}
The effects of our individual modifications for black hole growth and
the final success of the VEDISH model can also be explicitly seen in
Figs. \ref{QLF2} and \ref{QLF}, where the bolometric AGN luminosity
function (AGN LF) is plotted at different redshifts, and again
compared with the observational compilation from \citet{Hopkins07}.
Fig. \ref{QLF2} shows that the DI model raises the low-luminosity end
by about one order of magnitude in AGN number densities, while the VE
models lowers the high-luminosity end by more than two orders of
magnitude at $z \leq 1.5$. The SH model changes the AGN number
densities at $z \geq 3$ by lowering the low-luminosity end and raising
the high-luminosity end, each of them by about one order of magnitude.
The cumulative effect of the separate alterations is shown by the
DISH and the VEDISH model (Fig. \ref{QLF}). While the DISH model still
fails to reproduce the high luminosity end at low redshift, the
VEDISH model represents a fairly good match with the observational
data for the whole redshift range.

\subsection{X-ray Luminosity Functions}\label{Xlum} 

We now compare our model predictions with recent observational
determinations of the AGN LF in the hard and soft X-ray bands
(\citealp{Hasinger05, Ebrero09, Aird10, Fiore12}). In contrast to the
previous section, we do not attempt to correct the observations for
obscuration, but instead apply an obscuration correction to our
models.  We convert the modeled, bolometric luminosities into hard and
soft X-ray luminosities ($0.5-2$ keV and $>2$ keV) using the
bolometric correction according to \citet{Marconi04}. In their study,
the hard and soft X-ray luminosities $L_{\mathrm{HXR}},
L_{\mathrm{SXR}}$ are approximated by the following third-degree
polynomial fits:
\begin{eqnarray}
\log(L_{\mathrm{HXR}}/L_{\mathrm{bol}}) = -1.54 - 0.24\mathcal{L} -
0.012 \mathcal{L}^2 + 0.0015 \mathcal{L}^3\\
\log(L_{\mathrm{SXR}}/L_{\mathrm{bol}}) = -1.65 - 0.22\mathcal{L} -
0.012 \mathcal{L}^2 + 0.0015 \mathcal{L}^3
\end{eqnarray}
with $\mathcal{L} = \log(L_{\mathrm{bol}}/L_{\odot}) - 12$.  These
corrections are derived from template spectra, which are truncated at
$\lambda > 1\ \mu m$ in order to remove the IR bump and which are
assumed to be independent of redshift (therefore the resulting
bolometric corrections are also assumed to be redshift independent).
Additionally, we apply a correction for obscuration to the model
luminosities, as suggested by several observational studies
(\citealp{Ueda03, Hasinger04, LaFranca05}), in which it has been shown
that the fraction of obscured AGN is luminosity dependent and
decreases with increasing luminosity. While older studies such as
\citet{Ueda03} and \citet{Steffen03} did not find a clear dependence
of obscuration on redshift, several recent observational studies
(\citealp{Ballantyne06, Gilli07, Hasinger08}) propose a strong
evolution of the obscured AGN population (with the relative fraction
of obscured AGN increasing with increasing redshift). Here, we follow
the study of \citet{Hasinger08}, where they compare the same AGN in
both the soft and hard X-ray band so that they can derive an
approximation for the obscured fraction in the soft X-ray band. The
obscured fraction is then given by this equation:
\begin{eqnarray}
f_{\mathrm{obsc}} = -0.281 (\log(L_{\mathrm{LXR}})-43.5) + 0.279(1+z)^{\alpha}.
\end{eqnarray}
where they find that a value of $\alpha = 0.62$ provides the best fit
to their observational data.  By calculating the obscured fraction
of AGN in the soft X-ray band, we can model the visible fraction of
AGN $f_{\mathrm{vis}} = 1-f_{\mathrm{obsc}}$ and thus, the visible
number density of AGN in the soft X-ray range is given by:
\begin{eqnarray}
\Phi_{\mathrm{vis}}(L_{\mathrm{SXR}}) = f_{\mathrm{vis}} \times
\Phi_{\mathrm{total}}(L_{\mathrm{SXR}}) 
\end{eqnarray}

Fig. \ref{SXR} shows the \textit{soft} X-ray luminosity function for
different redshift ranges predicted by our FID, DISH and VEDISH
models, compared with observations from \citet{Hasinger05} and
\citet{Ebrero09}. We show the model predictions with and without the
obscuration correction described above. For the high-luminosity end
($L_{\mathrm{SXR}} > 10^{45}\ \mathrm{erg/s}$), obscuration does not
influence our results and thus, one can see the same trends as already
discussed in Fig. \ref{QLF}.  Turning to the low-luminosity end at low
redshifts ($z \leq 0.8$), the FID model matches the observational
data, while the DISH and VEDISH model overproduce the number density
of AGN. However, considering obscuration effects leads to better
agreement of the VEDISH model with the observed number densities than
the FID model. At high redshift ($z>0.8$), both the FID and the VEDISH
model over-predict the number of moderately luminous AGN by almost the
same amount, and both models including obscuration can achieve a
fairly good agreement with the observations. Overall, the VEDISH model
including AGN obscuration can predict the observed soft X-ray
luminosity function reasonably well, supporting the adoption of a
redshift-dependent obscured fraction.  
\citet{Fanidakis10} use the
same bolometric conversion for calculating soft X-ray luminosities and
the same dust obscuration correction as in our study. They show that
they are able to match the soft X-ray luminosity functions from the
observational study of \citet{Hasinger05} as well.

Fig. \ref{HXR} illustrates the \textit{hard} X-ray luminosity
functions predicted by our models for different redshift ranges,
compared with observational data (\citealp{Ebrero09},
\citealp{Aird10}, \citealp{Fiore12}, \citealp{Fiore12}). The VEDISH
model is able to reproduce the high-luminosity end pretty well,
whereas it over-predicts the number of AGN at the low-luminosity end at 
all redshifts (with a larger discrepancy at higher redshift). One
possible explanation might be again due to obscuration as even
$2-10\ \mathrm{keV}$ X-ray surveys might miss a significant fraction
of moderately obscured AGN ($~ 25$\% at $N_H =
10^{23}\ \mathrm{cm}^{-2}$) and nearly all Compton-thick AGN ($N_H >
10^{24}\ \mathrm{cm}^{-2}$, \citealp{Treister04, Ballantyne06}). From
fits to the cosmic X-ray background, \citet{Gilli07} predict that both
moderately obscured and Compton-thick AGN are as numerous as
unobscured AGN at luminosities higher than $\log(L_{0.5-2{\rm keV}})> 43.5
[\mathrm{ergs/s}]$, and four times as numerous as unobscured AGN at
lower luminosities ($\log(L_{0.5-2 {\rm keV}}) < 43.5 [\mathrm{ergs/s}]$).
For this reason, we made the very simplified assumption that the
fraction of obscured AGN in the hard X-ray band is the same as in the
soft X-ray band.  This results in a fairly good match to the observed
low-luminosity end at all redshifts, except around $z \sim 1$ where
the obscuration-corrected model under-predicts the number of faint
AGN. Therefore, our model results are consistent with the existence of
an obscured fraction of AGN in the hard X-ray band (Compton-thick AGN)
that is of the same order of magnitude as the obscured fraction in the
soft X-ray band.  Interestingly, \citet{Fanidakis10} show that they
are able to match the hard X-ray luminosity functions from an
observational study of \citet{Ueda03} \textit{without} including any
obscuration effects. However, their predictions are only illustrated
for a comparatively small redshift range of $0.2<z<1.6$ and for hard
X-ray luminosities larger than $L_{\mathrm{HXR}} >
10^{42}\ \mathrm{erg/s}$, where our unobscured VEDISH model
predictions are also in good agreement with the observational data.

\section{Eddington ratio distributions}\label{Eddratioevol} 

\begin{figure}
\begin{center}
  \epsfig{file=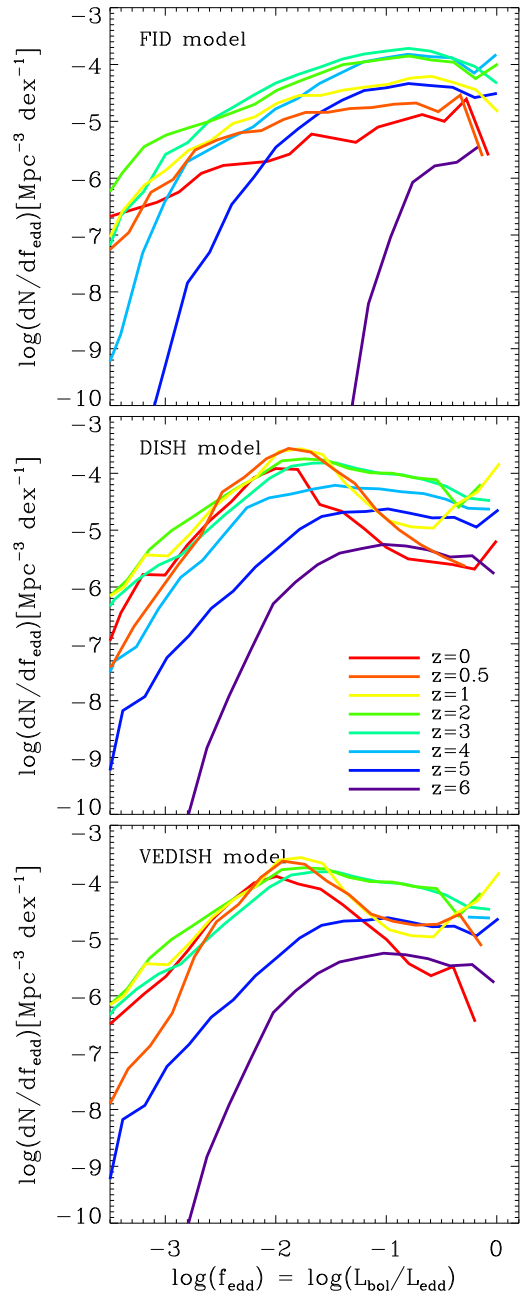, width=0.4\textwidth}
  \caption{Eddington ratio distributions for the FID, DISH and the
    VEDISH models (upper, middle and lower panel panel, respectively).
    Different colors correspond to different redshifts
    ($z=0.5,1,2,3,4,5,6$). Note that here we show only accreting
    black holes with bolometric luminosities larger than
    $10^{43}\ \mathrm{erg/s}$. The modifications in the VEDISH model
    cause the peaks of the distributions to shift towards smaller
    Eddington ratios with time, in qualitative agreement with
    observational studies (\citealp{Vestergaard03, Kollmeier06,
      Kelly10, Schulze10})} {\label{Eddratio}}
\end{center}
\end{figure}

Having assessed our models by comparing with the observed AGN LF, we
now examine the consequences of our modifications of the black hole
growth prescriptions on the Eddington ratio distribution and on the
AGN luminosity-black hole mass plane. In Fig. \ref{Eddratio} we show
the redshift evolution of the Eddington ratio distributions for the
FID (upper panel), the DISH (middle panel) and the VEDISH model (lower
panel). Different colors indicate different redshift steps
($z=0,0.5,1,2,3,4,5,6$). Consistent with our previous investigation,
and typical limits of observational samples, we consider only AGN with
bolometric luminosities larger than $10^{43}\ \mathrm{erg/s}$. In all
models, the fraction of AGN that are are accreting at smaller
Eddington ratios increases strongly with decreasing redshift. This is
because at later times, black holes spend less of their time in the
first regime, i.e. accreting at the Eddington rate, but mainly reside
in the second regime, the power-law decline dominated ``blowout''
accretion phase. These black holes are relics from an earlier, more
active phase with higher accretion rates. While, however, in the FID
model the Eddington ratios are still peaking between
$f_{\mathrm{edd}}=0.1-1$ at \textit{all} redshifts, in the DISH and
the VEDISH models the peaks of the distribution curves are clearly
shifted towards smaller Eddington ratios with decreasing
redshift. This is mainly caused by the additional black hole accretion
due to secular evolution processes. At $z=6$ the distributions of the
DISH and VEDISH models peak at $f_{\mathrm{edd}} \approx 0.1$, while
at $z=0$ the peaks are located around $f_{\mathrm{edd}} \approx
0.01$. Thus, the majority of AGN at $z=0$ are not radiating at or near
the Eddington limit anymore, which is in qualitative agreement with
observational studies (\citealp{Vestergaard03, Kollmeier06, Kelly10,
  Schulze10}). For example, \citealp{Kelly10} find that the
Eddington ratio distribution (using broad-line quasars between
$z=1-4$) peaks at an Eddington ratio of $f_{\mathrm{edd}} =0.05$.
Furthermore, the large seed black hole masses and the large scatter in
the accreted gas mass in the DISH and VEDISH models lead to a
significantly larger number of accreting black holes with Eddington
ratios between $0.01<f_{\mathrm{edd}}<1$ at $z>4$ than in the FID
model. Finally, the VEDISH model additionally shows that as a
consequence of the sub-Eddington accretion rate limit (dependent on
the cold gas fraction), the number of black holes accreting close to
the Eddington limit at $z=0$ is reduced by about one order of
magnitude compared to the DISH model. Altogether, we find that the
downsizing behavior seems to imply that the peaks of the Eddington
ratio distributions are shifted towards smaller Eddington ratios with
decreasing redshift, and the number of black holes accreting close to
the Eddington-rate is low at $z=0$ ($\log(dN/df_{\mathrm{edd}}) =
-6.5\ \mathrm{Mpc}^{-3}\ \mathrm{dex}^{-1}$), while at high redshifts
a large number of black holes ($\log(dN/df_{\mathrm{edd}}) =
-5.2\ \mathrm{Mpc}^{-3}\ \mathrm{dex}^{-1}$) are accreting within a
broad range of Eddington ratios $0.01<f_{\mathrm{edd}}<1$.

\section{Luminosity-black hole mass-plane}\label{BHlumevol} 

\begin{figure*}
\begin{center}
  \epsfig{file=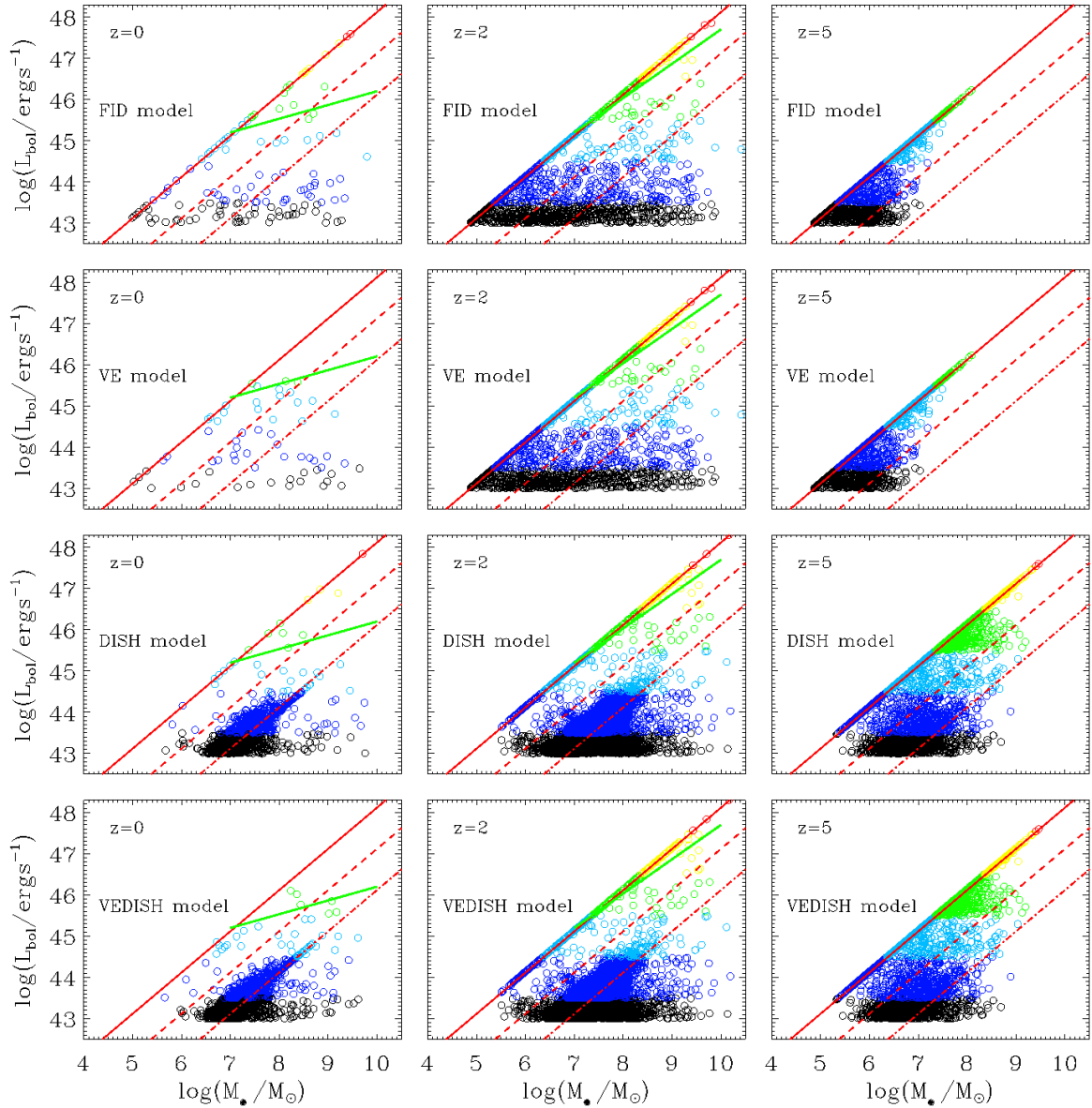, width=0.9\textwidth}
  \caption{Bolometric luminosity versus black hole mass at $z=0$ (left
    column), $z=2$ (middle column) and $z=5$ (right column). Different
    rows correspond to the FID (first row), the VE (second row), the
    DISH (third row) and the VEDISH (fourth row) model. The open
    circles illustrate actively accreting black holes at each
    redshift. The different colors correspond to different bolometric
    luminosity bins as defined in Fig. \ref{Numdens_all}. The red
    solid lines always illustrate accretion at the Eddington rate
    $f_{\mathrm{edd}} = 1$, while the red dashed and dotted dashed
    lines depict Eddington ratios of $f_{\mathrm{edd}} = 0.1$ and
    $f_{\mathrm{edd}} = 0.01$, respectively. The green solid lines
    show the sub-Eddington limit from \citet{Steinhardt10}. We find
    that the observed upper envelope in Eddington ratio is reproduced
    well by our adopted gas fraction dependent accretion limit. }
          {\label{Lum_bh}}
\end{center}
\end{figure*}

Fig. \ref{Lum_bh} shows the bolometric AGN luminosity-black hole
mass-plane at $z=0$ (left column), $z=2$ (middle column) and $z=5$
(right column) for different SAMs (first row: FID model, second row:
VE model, third row: DISH model and fourth row: VEDISH model). The
circles in each panel correspond to our model AGN with a bolometric
luminosity cut of $L_{\mathrm{bol}} > 10^{43}\ \mathrm{erg/s}$ and
they are color coded according to their bolometric luminosity as
defined in Fig. \ref{Numdens_all}. We also show lines indicating
accretion at the Eddington rate, an Eddington ratio of
$f_{\mathrm{edd}} = 0.1$ and an Eddington ratio of $f_{\mathrm{edd}} =
0.01$. In addition, we show the observational limit in the
$L_{\mathrm{bol}}-M_{\bullet}$-plane according to
\citet{Steinhardt10}.

At $z=5$ (right column of Fig. \ref{Lum_bh}), in the FID and VE
models, black holes within a mass range of $10^{4.7}<M_{\bullet}<10^8
M_{\odot}$ are active, where the luminosity is almost linearly
correlated with black hole mass, indicating accretion at or very close
to the Eddington rate. As a consequence of the heavy seeding mechnism
and of the halo mass limit for black hole formation, the mass range of
active black holes is shifted towards larger values
$10^{5.5}<M_{\bullet}<10^{9.5} M_{\odot}$ in the DISH and the VEDISH
models. As we showed previously, in order to reproduce the observed
downsizing trend these bright AGN
($L_{\mathrm{bol}}>10^{46.5}\ \mathrm{erg/s}$) need to appear already
at these early times.  Turning to $z=2$ (middle column of
Fig. \ref{Lum_bh}), we see that in all models the number of AGN with
massive BH ($M_{\bullet} > 10^{9}M_{\odot}$) has increased relative to
$z=5$.  Compared to the accretion limit of \citet{Steinhardt10}, all
of our model results are in acceptable agreement, although we predict
some massive BH accreting at or very close to the Eddington rate which
strictly speaking are not `allowed' according to the
\citet{Steinhardt10} results. Also for all models, the relation
between black hole mass and bolometric luminosity becomes much broader
--- black holes with masses $M_{\bullet} \approx 10^8 M_{\odot}$ can
now also power moderately luminous AGN with $L_{\mathrm{bol}} \approx
10^{43} \mathrm{erg/s}$ as they are accreting with Eddington ratios
below $f_{\mathrm{edd}} < 0.01$. The probability for black holes with
$M_{\bullet} > 10^7 M_{\odot}$ to accrete at Eddington ratios below
$f_{\mathrm{edd}} = 0.01$ is even higher than to accrete at larger
Eddington ratios. This is due to the power-law decline accretion phase
that black holes are experiencing: massive black holes powering
moderately luminous AGN are remnants of former, high-luminous AGN. In
the DISH and VEDISH models, we can additionally see the effect of
black hole accretion due to disk instabilities: the number of AGN with
black hole masses of $10^7 M_{\odot} < M_{\bullet} < 10^8 M_{\odot}$
and with luminosities between
$10^{43}<L_{\mathrm{bol}}<10^{45}\ \mathrm{erg/s}$ is significantly
increased, accreting with Eddington ratios around
$f_{\mathrm{edd}}\approx 0.01$. As Seyfert galaxies are mainly spiral
galaxies with black hole masses in the range of $ 10^7 M_{\odot} <
M_{\bullet} < 10^8 M_{\odot}$ and are --- compared to quasars --- only
moderately luminous, disk instabilities seem indeed to provide one of
the most important trigger mechanisms for their nuclear
activity. Moreover, in the DISH and VEDISH models, no black holes
below $10^{5.5} M_{\odot}$ are active, in contrast to the FID and the
VE models. This is due to the halo mass limit for black hole
formation.

Finally, at redshift $z=0$ (left column of Fig. \ref{Lum_bh}), the
number of actively accreting black holes is significantly reduced
compared to $z=2$. The DISH and the VEDISH models show again the
effect of the additional accretion channel due to disk instabilities,
increasing the number of moderately luminous AGN, with typical black
hole masses of Seyfert galaxies ($ 10^7 M_{\odot} < M_{\bullet} < 10^8
M_{\odot}$). While, however, in the FID and the DISH models, bright
AGN can exist with massive black holes accreting at and close to the
Eddington rate, in the VE and VEDISH models the limited accretion
regulated by the cold gas fraction suppresses the appearance of these
bright AGN. Compared to the observed accretion limit of
\citet{Steinhardt10}, our results in the VE and VEDISH model are in
very good agreement with the observed limit. This strongly suggests
that the dependence of the accretion rates on the cold gas content
might provide a possible physical origin for the observed accretion
limit and thus, the reduced number of bright AGN at low redshifts.

Interestingly, in the study of \citet{Fanidakis10}, they obtain almost
no evolution of their luminosity-black hole mass relation within a
redshift range $0.5<z<2$. Black holes above $M_{\bullet} > 10^9
M_{\odot}$ always accrete below $f_{\mathrm{edd}} = 0.01$ (ADAF
regime), as in our VE and VEDISH models. However, in clear contrast to
this work, in this redshift range, they still have super-Eddington
accretion, in particular for black hole masses between $10^8 M_{\odot}
< M_{\bullet} < 10^9 M_{\odot}$ (resulting in $L_{\mathrm{bol}} >
10^{46}\ \mathrm{erg/s}$). This seems to be, however, in contrast to
several observational studies, such as \citet{Steinhardt10}, which do
not find black holes with masses $M_{\bullet} > 10^7 M_{\odot}$
accreting near or at the Eddington limit at low redshifts $z=0.2-0.4$.

\section{Summary and discussion}\label{downdis}

In this study, we have used the semi-analytic model of S08 to study
the origin of the observed anti-hierarchical behavior of BH activity
within the framework of hierarchical structure formation. In the
original S08 SAM (FID), all AGN are merger-driven, with the
prescription for BH accretion and the light curve models based on
hydrodynamic binary merger simulations. In these models, BH growth is
self-regulated, resulting in a tight scaling relation between BH mass
and spheroid mass. Following a merger, BH are allowed to grow until
they reach a critical mass, at which the luminosity emitted by the AGN
is sufficient to power a pressure-driven outflow that is able to slow
down subsequent accretion, and eventually to unbind the gas in the
galaxy. In this picture, BH accrete at the Eddington limit until they
reach this critical mass, after which, in the ``blowout'' phase, the
accretion rate declines as a power law function of time. The FID model
reproduces the observed number densities of AGN at $z\sim2$ over a
wide range of bolometric luminosity, but in its original form it does
not reproduce the observed downsizing trend: at low redshift, the FID
model overproduces the number of bright AGN and under-predicts the
number of moderately luminous AGN, while at high redshift, this trend
is reversed. Therefore, we extended the FID model by considering
different modifications to the physical recipes for black hole growth,
and investigated their effect on the AGN evolution in different
luminosity bins. We summarize our main findings:
\begin{enumerate}[1.]
\item{\textbf{A sub-Eddington limit dependent on the cold gas fraction
    of the host galaxy (VE model)}: The FID model overproduces
  luminous AGN at low redshift ($z<1$). We found that introducing a
  sub-Eddington cap on the BH accretion rate which was dependent on
  the gas fraction of the progenitor galaxies resulted in improved
  agreement with the observations in this regime. In our model, the
  gas content of massive galaxies is strongly dependent on the Radio
  Mode AGN feedback, which suppresses cooling in massive halos at low
  redshift. Low cold gas fractions may retard the loss of angular
  momentum due to smaller viscosity and thus, the cold gas flow onto
  the central black hole may be suppressed. However, we found that we
  only obtained a good match to observations when we implemented this
  accretion rate cap at $z \leq 1$. Extending the same gas-fraction
  dependent limit to all redshifts results in too few of the brightest
  AGN at redshifts between $1<z<4$.  One possible explanation might be
  that the evolution of the cold gas content in the FID SAM may not be
  correctly modeled. This could be due to incorrect or simplified
  cooling recipes \citet{Hirschmann11,Lu11} or inadequate star
  formation recipes (Caviglia \& Somerville, in prep).}
\item{\textbf{An additional BH accretion triggering mechanism due to disk
    instabilities (DI model)}:
The second main result of our study is that gas accretion
onto black holes due to disk instabilities seems to be a
non-negligible trigger mechanism for moderately luminous
AGN with black hole masses $10^7 M_{\odot} < M_{\bullet} < 10^8
M_{\odot}$ at low redshift. This increases the number of AGN with
bolometric luminosities $ 10^{43}\ \mathrm{erg/s} < L_{\mathrm{bol}} <
10^{45}\ \mathrm{erg/s}$, i.e. Seyfert galaxies, by about one order of
magnitude. 
Our results therefore favor a `hybrid' picture in which major merger events are
the main driver of luminous AGN, especially at high redshift, while
disk instabilities are the main mechanism powering moderately luminous
Seyfert galaxies at low redshift (consistent with the picture
suggested by e.g. \citealp{Hopkins09}).
Note that the studies based on the \textsc{Munich} and the
\textsc{Galform} model have come to rather different conclusions (see
discussion below).
\item{\textbf{Heavy seed black holes with a halo mass threshold for
    seed formation (SH model)}}: Our FID model does not produce
  enough very luminous QSOs at high redshift ($z>5$). 
Our third, main conclusion is that a heavy seeding mechanism for black
holes provides a physically motivated way to increase the predicted
number density of very luminous QSOs at high redshift, in line with
observations. This is in agreement with the studies of
\citet{Volonteri08} and \citet{Volonteri10}, which showed that either
massive black hole seeds are required or less massive seeds have to
accrete at super-Eddington rates. Unfortunately, current observational
constraints are not sufficient to favor one of these possibilities.
This may be possible with the next generation of planned X-ray
missions (e.g. WFXT, IXO). As an implicit consequence of massive black
hole seeds, we have also assumed that seeds are not able to form in
dark matter halos below $2\times10^{11}\ M_{\odot}$, as cold gas might
not be able to collapse in these halos due to their shallow potential
wells \citep{Volonteri08}. This simultaneously reduces the number of
faint AGN at high redshift, in better agreement with observations.  }
\end{enumerate}

We find that the FID model with a combination of the above
modifications (=VEDISH model) can reproduce the downsizing trend
fairly well and is e.g. able to match the observational compilation
of \citet{Hopkins07}. 
Moreover, we have shown that the additional modifications do not
change basic galaxy and black hole properties at $z=0$ significantly
and thus, our model galaxies are still in agreement with the observed
local stellar mass function, black hole mass function and the black
hole-bulge mass relation. However, as in other semi-analytic models we
find that at high redshift ($z \gtrsim 1$) the number of low-mass
galaxies at high redshift is overestimated and the number of high-mass
galaxies may be underestimated. These predictions are insensitive to
our modifications to the black hole seeding and growth recipes, but of
course our predictions for AGN number densities are impacted by these
discrepancies in correctly predicting the galaxy population.

For the ``best-fit'' VEDISH model we find that the peaks of the
distributions of Eddington ratios move towards smaller values with
decreasing redshift in qualitative agreement with observational
studies (\citealp{Vestergaard03, Kollmeier06, Netzer07, Kelly10,
  Schulze10}). Additionally, the results of the VEDISH model are in
excellent agreement with the observed accretion limit in the
$L_{\mathrm{bol}}-M_{\bullet}$-plane of \citet{Steinhardt10}. In our
model, this behavior results from the gas-fraction dependent accretion
rate cap that we applied. 

Despite the success of the final VEDISH model in reproducing the
bolometric AGN luminosity function from the Hopkins compilation, the
number of faint AGN still appears to be overestimated at high
redshifts $4>z>2$. One possible explanation for this discrepancy is a
redshift-dependent dust obscuration, which was neglected in
\citet{Hopkins07}. Therefore, we have compared our model results
directly to the observed AGN LF in the soft- and hard X-ray band by
assuming the obscuration model from \citet{Hasinger08} for the soft-
and also for the hard X-ray luminosities, as even $2-10\ \mathrm{keV}$
X-ray surveys will miss a significant fraction of moderately obscured
AGN and nearly all Compton-thick AGN (\citealp{Treister04,
  Ballantyne06}).  With our VEDISH model including obscuration effects
we achieve excellent agreement with the observed AGN LF in the soft
and hard X-ray bands for the whole luminosity range at \textit{all}
redshifts.  Therefore, our results suggest that the obscured and
Compton-thick AGN missed in deep X-ray surveys likely constitute a
significant fraction of the total AGN population at all luminosities,
in particular of moderately luminous AGN at high redshift. However, we
note that although obscuration effects contribute to the late peak in
the number densities of moderately luminous AGN, in our model they
\textit{cannot} account for the observed downsizing trend completely
(see FID model and dust obscuration in Figs. \ref{HXR} and \ref{SXR}).

Comparing our conclusions with those of some previous studies (see
Section \ref{previous}), we summarize the following main points:
\citet{Marulli08} found that a model based on the original
\textsc{Munich} semi-analytic code and BH growth recipe
\citep{Kauffmann00,Croton06}, and with a light curve model similar to
the one adopted here, underproduced luminous AGN at high redshift and
slightly overproduced them at low redshift ($z\sim 0$), and
underproduced low-luminosity AGN at low redshift ($z\lesssim1$). As a
result, they introduced an ad-hoc scaling of the black hole accretion
efficiency with an explicit redshift dependence. With this ``best
fit'' model, they were able to reproduce the bolometric AGN luminosity
function fairly well except at very high redshift ($z>4$). Thus, their
results are qualitatively very consistent with ours --- in effect,
they found that the same qualitative modifications to the model were
necessary, and achieved this via an explicit modification of the
accretion efficiency, while we have tried to achieve this by adding
more physically motivated effects. However, contrary to our study,
they find that they do not need AGN activity triggered by disk
instabilities, but can explain all AGN activity with a merger model.

In contrast, in the study based on the Durham \textsc{Galform} model
(\citealp{Fanidakis10}), disk instabilities are the major triggering
mechanism for black hole activity for \textit{all} luminosities at
\textit{all} redshifts. With their model they can reproduce the AGN
luminosity function at all redshifts: the low-luminosity end of the
AGN luminosity function is mainly due to their ADAF model (cold gas
accretion onto the black hole in a hot halo), whereas in our model it
is due to accretion in the power-law decline regime as well as black
hole accretion from disk instabilities. At high redshift, they predict
that super-Eddington accretion is responsible for producing luminous
AGN. At this point, it is not clear whether a heavy seeding mechanism
(as assumed in this study) or super-Eddington accretion describes the
correct physical processes.  For that question, the improvement of
high-redshift AGN observations will be of crucial importance, both by
enlarging the current high-z AGN samples and by reducing the current
uncertainty originating from incompleteness problems. They mainly
attribute the downsizing trend to dust obscuration effects.

Thus the main triggering mechanism of most of the black hole growth in
the universe remains a major open question. It is clear that with the
allowed freedoms from various modeling and observational
uncertainties, it is possible to reproduce the \emph{number densities}
of AGN (luminosity functions) as a function of redshift within a range
of scenarios, from a pure merger scenario, to a hybrid merger+disk
instability scenario such as the one we have suggested, to a pure
disk-instability driven scenario. Hopefully, this issue will be
clarified by further studies of the morphology of AGN hosts, now
possible out to the peak of luminous AGN activity $z\sim 2$ with Wide
Field Camera 3 (WFC3) on Hubble
(e.g. \citealp{Schawinski11,Kocevski12,Rosario11}), and the enhancement
of AGN activity in close pairs over a range of redshifts and
luminosities. We intend to make a more quantitative comparison of our
predicted host properties with observations in a future work.

Overall, we conclude that the models presented here provide a
plausible attempt to understand the complex scenario of black hole and
galaxy co-evolution, and to predict the downsizing trend within the
framework of hierarchical clustering. However, there still remain many
uncertainties in modelling the formation and evolution of black holes,
which hopefully ongoing and future observational facilities will help
to constrain.

\section*{Acknowledgments}
This research was supported by the DFG Cluster of Excellence `Origin
and structure of the universe'. MH acknowledges financial support from
the European Research Council under the European Community's Seventh
Framework Programme (FP7/2007-2013)/ERC grant agreement n. 202781.  We
thank Francesco Shankar, James Aird, Jacobo Ebrero and Fabrizio Fiore
for providing us with observational data and David Alexander, Fabrizio
Fiore, Francesco Shankar, Phil Hopkins, Brant Robertson, TJ Cox, Lars
Hernquist, and Yuexing Li for fruitful discussions.

\bibliographystyle{mn2e}
\bibliography{Literaturdatenbank}

\label{lastpage}

\end{document}